\documentclass[review]{elsarticle}
\usepackage{geometry}
\usepackage[utf8]{inputenc}
\usepackage{mathrsfs,amssymb,amsmath}
\usepackage{url}
\usepackage{graphicx}

\usepackage{lineno,hyperref}

\DeclareMathAlphabet{\mathbfsf}{\encodingdefault}{\sfdefault}{bx}{n}
\newcommand{\twoform}[1]{\mathbfsf{#1}}

\journal{Journal of \LaTeX\ Templates}

\begin{document}

\begin{frontmatter}

\title{Explicit physics-informed neural networks for non-linear upscaling closure: the case of transport in tissues}

\author[OSU]{Ehsan Taghizadeh  \fnref{1}}
\author[Oxford]{Helen M. Byrne \fnref{2}}
\author[OSU]{Brian D. Wood     \fnref{3}}

\address[OSU]{Chemical, Biological, and Environmental Engineering, Oregon State University, OR 97330, USA}
\address[Oxford]{Mathematical Institute, University of Oxford, Oxford OX2 6GG, UK}

\fntext[1]{ehsan.taghizade@gmail.com}
\fntext[2]{helen.byrne@maths.ox.ac.uk} 
\fntext[3]{Corresponding author: brian.wood@oregonstate.edu}

\begin{abstract}
In this work, we use a combination of formal upscaling and data-driven machine learning for explicitly closing a nonlinear transport and reaction process in a multiscale tissue.  The classical effectiveness factor model is used to formulate the macroscale reaction kinetics. We train a multilayer perceptron network using training data generated by direct numerical simulations over microscale examples. Once trained, the network is used for numerically solving the upscaled (coarse-grained) differential equation describing mass transport and reaction in two example tissues.  The network is described as being \emph{explicit} in the sense that the network is trained using macroscale concentrations and gradients of concentration as components of the feature space.

Network training and solutions to the macroscale transport equations were computed for two different tissues. The two tissue types (brain and liver) exhibit markedly different geometrical complexity and spatial scale (cell size and sample size).  The upscaled solutions for the average concentration are compared with numerical solutions derived from the microscale concentration fields by \emph{a posteriori} averaging.
There are two outcomes of this work of particular note: 1) we find that the trained network exhibits good generalizability, and it is able to predict the effectiveness factor with high fidelity for realistically-structured tissues despite the significantly different scale and geometry of the two example tissue types; and 2) the approach results in an upscaled PDE with an effectiveness factor that is predicted (implicitly) via the trained neural network. This latter result emphasizes our purposeful connection between conventional averaging methods with the use of machine learning for closure; this contrasts with some machine learning methods for upscaling where the exact form of the macroscale equation remains unknown.
\end{abstract}

\begin{keyword}
explicit physics-informed neural networks \sep deep learning \sep tissue transport \sep nonlinear kinetics \sep upscaling \sep effectiveness factor
\end{keyword}

\end{frontmatter}
\section{Introduction}

Developing closures associated with the upscaling of nonlinear continuum mechanical problems is an enduring challenge.  One solution to this problem, fostered by increased computational speed and capacity, has been the development of methods to directly resolve all relevant scales of the phenomena of interest.  While this has been applied effectively to the problem of, for example, the momentum balances that describe turbulence, the approach is still too cost prohibitive (in terms of computational requirements) to be used routinely for many problems of interest.  Thus, there is a continuing need for appropriately upscaled (or \emph{coarse-grained} or \emph{homogenized}) representations of nonlinear continuum mechanical problems.

Upscaling can be accomplished via a number of approaches ranging from formal averaging methods to various numerical schemes; a review of these methods (with an emphasis on nonlinear problems) has been reported in \citep{matouvs2017review, burzawa2020, peng2020multiscale}.  Regardless of the approach, 
the process of eliminating the microscale variables in coarse-grained problems is known generally as the \emph{closure problem}.  For nonlinear problems, there are no general methods for exact closures.  The use of machine learning (ML) methods represents a relatively new option for closing nonlinear problems.  With sufficient training data, ML methods (such as any of a number of neural networks types, support vector methods, etc.) have the ability to \emph{learn} how to represent such data.  

In this paper, we focus specifically on coupling a \emph{formal upscaling method} \citep{whitaker1999}, for conducting the coarse-graining of the problem, with the use of a deep learning approach, for effecting the nonlinear closure.  For this problem, upscaling the transport and reaction problem in biological tissues is investigated by starting at the microscale level of representation of the quantities of interest (QoIs).  We adopt a simple two-phase description of the system with classical hyperbolic kinetics representing the reaction term.  Because the reaction term is nonlinear, there is no formal exact scheme for closing the problem.  We investigate the use of a deep multilayer perceptron (MLP) network for learning the an \emph{effectiveness factor} representation of the reaction term with high fidelity. More importantly, the trained neural network shows high generalizability even for problems with very different scales and geometry than the training data examples, implying that the learned mapping function can be used in a wide range of real applications.  Physical constraints for the problem are imposed by introducing the source term in the features space in order to retain the mass balance.

\subsection{Background}

Machine learning (ML) is a discipline that has seen a renaissance in applications, mostly due to a combination of increasing processor speeds, computational and storage capacities, and the development of new algorithms.   While there are still many questions regarding the interpretability of these methods \citep{molnar2020}, they provide one of the few methods that have offered solution options for complex problems, in particular nonlinear ones.

There has been some success in both learning and predicting the complex spatio-temporal behavior of linear and nonlinear partial differential equations (PDEs) \citep{ranade2020discretizationnet, zhuang2020learned,rudy2017data,raissi2018hidden,raissi2019physics}. A subset of this work has focused on discovering coarse-grained representations of the PDEs under consideration using ML methods \citep{bar2019learning}.  These approaches fall primarily into two major categories.

\begin{enumerate}
    \item Methods for directly \emph{learning} the appropriate coarse-grained PDE.  The number of different approaches pursued here is vast.  It includes the use of simple neural networks \citep{pawar2020interface}, recurrent neural networks \citep{arbabi2020linking}, deep neural networks with dimension reduction via diffusion maps \citep{lee2020coarse}, recurrent neural network architectures with dimension reduction accomplished via an autoencoder scheme \cite{vlachas2020learning}, multilayer perceptrons for generating reduced-order metamodels \citep{burzawa2020}, and direct / constrained equation learning methods \citep{bakarji2021data}.  Because these models are designed to learn the appropriate macroscale behavior from an appropriate suite of ML approaches, the end result is frequently high-fidelity predictive ability from the generated network, but usually without the generation of an explicit macroscale equation for the process.\\
    
    \item Methods in which the macroscale form of the PDE is specified (via any of a number of formal methods for upscaling), and then the problem is \emph{closed} by learning from a large number of microscale examples.  There are occasions where the development of an explicitly-defined macroscale balance equation is desirable.  There are a number of classical upscaling methods that have been used to accurately average microscale balance equations to generate macroscale representations in a mathematically formal and robust way (reviews of such approaches can be found in \citep{davit2013homogenization,battiato2019theory}).  These approaches often employ use physical information uncovered from the upscaling process (e.g., solution invariant, bounds for the variable scale, smoothness requirements, or the specification of source terms in the of the closure problem) to constrain the problem.   This approach has recently had good success in predicting nonlinear closures for turbulence \citep{ling2016,duraisamy2019}.  
\end{enumerate}
In this work, we adopt the second of these two approaches, with an application to the coarse-graining of chemical transport and reaction in biological tissues.

It is important to stress that learning methods are not, in general, mere fitting algorithms.  The concept of \emph{learning} includes both 1) the ability to fit a function to a high-dimensional data set, and 2) a demonstrated ability for the learned model to have high predictive fidelity for \emph{new data} that were not part of the training (i.e., the model \emph{generalizes} well).  

\subsection{Objectives and Outline}

For continuum mechanical descriptions of biological tissues, microscale representations for describing both physical and biological processes have become significantly more widespread  \citep{wang2020impact}.  While sub-cell-level representations of the biophysical and biochemical processes in tissues has several advantages (e.g., allowing direct representation of cells with different phenotypes), such representations suffer from the same problems as those described for turbulence.  For many practical applications, both the number of degrees of freedom and the presence of nonlinearities make the direct computation of the continuum mechanics at the microscale level impractical \citep{burzawa2020}. Hence, upscaling methods have been developed and employed to allow resolution in spatial and temporal domains where it is needed, while providing accurate but more economical methods for domains where resolution is not a priority.

In this paper, we develop a data-driven framework which combines upscaling, microscale numerical solutions, and feed forward neural networks to provide closure for the problem of transport and nonlinear reactions in tissues. Note that the proposed approach can be employed for all types of transport questions as long as training dataset is computationally feasible. Our approach is based on using the classical  \emph{effectiveness factor} \citep{thiele1939relation, truskey2004transport,shuler1972diffusive} to define the reaction rate in terms of average concentrations.  In short, the effectiveness factor can be defined as follows. Consider a nonlinear kinetic reaction rate $R = R(c)$ which depends upon the concentration, $c$.  A spatial average of the reaction rate (where $\langle \cdot \rangle$ represents the spatial averaging operation) is then specified by
\begin{equation}
 \langle R(c) \rangle = \eta(\boldsymbol{x}) R(\langle c \rangle)
 \label{effectiveness1}
\end{equation}
where $\boldsymbol{x}$ is a vector of descriptive model parameters determined from the physics of the problem. Here, $\eta(\boldsymbol{x})$ is a (potentially highly) nonlinear correction factor that accounts for the fact that the local concentrations, at the cell surface, may be significantly different from the average concentration, especially when intercellular reactions are present.  The effectiveness factor is then learned by using a feed-forward multi-layer perceptron (MLP) network to parameterize a least-squares fit of a training data set containing thousands of examples of features and $\eta(\boldsymbol{x})$ set.  In short, we seek to approximate the function $\eta$ by the following schematic compositional form, indicating $M$ sequential transformations of an initial linear combination, $l^{(0)}$,  of the feature space $\boldsymbol{x}$ (cf., \citep{kharazmi2019variational})

\begin{equation}
    \eta(\boldsymbol{x})\approx \hat{\eta}(\boldsymbol{x}) = \ell^{(f)}\circ T^{(M)}\circ \ldots \circ T^{(1)} \circ \ell^{(0)}
\end{equation}
Each of the $M$ transformations, $T^{(i)}$, is accomplished using any of a number of possible nonlinear activation (or basis) functions; conventionally these transformation layers are called \emph{hidden layers}.  For each such layer $T^{(i)}$, there are $N^{(i)}$ neurons, which may vary from layer to layer depending upon the network structure.  The output of each hidden layer is the input for the subsequent layer; usually the input to subsequent layers are constructed as simple linear combinations of the output from the previous layer, but other mappings are possible.  The final output is again generated by a linear combination from the output, $\ell^{(f)}$, of the last hidden layer.  Each layer will generally include one neuron of the $N^{(i)}$ that represents a constant (bias) term.

\begin{figure*}[t]
\begin{center}
\includegraphics[width=0.95\textwidth]{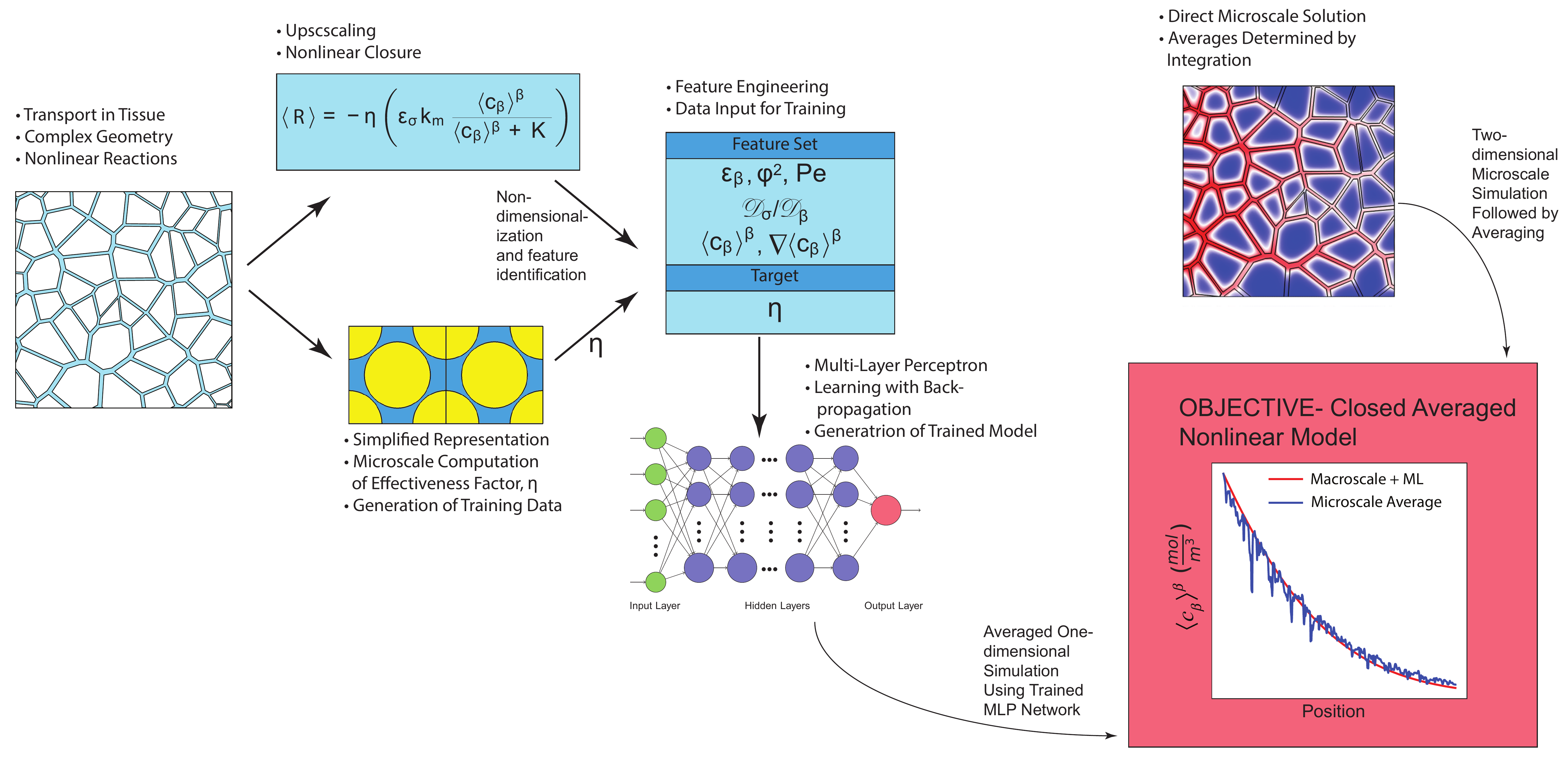}
\caption{Graphical representation for the upscaling and closure of transport and nonlinear reactions in biological tissues.}
\label{workflow}
\end{center}
\end{figure*}

A graphical presentation of the workflow for this process is provided in Fig.~\ref{workflow}.  Our approach has many similarities with the closure scheme proposed by \citep{kevrekidis2003}, but with an extension to neural networks for parameterizing the proposed closure and more intensive sampling of the microscale solutions. 

The remainder of the paper is organized as follows.  In \S 1, we outline the sub-cell-scale continuum-mechanical description of mass and momentum transport in a multiphase (intracellular phase, extracellular phase, and cell membrane) tissue. In \S 2, we summarize previous work in which the microscale problem has been upscaled using the method of volume averaging, introduce the definition of the effectiveness factor, and explain how it can be computed from direct microscale simulation on a representative volume. In \S 3, we describe the upscaled tissue transport model; most of the results in this section reflect results summarized from previous work by our group.  In \S 4, we describe the identification of the feature sets for learning, emphasizing how these features are selected using physically-driven assessments of the problem. In \S5 the feed forward neural network is described, and the process of developing the data sets and training the network are explored in detail.  In \S 6, validation of the trained model for the effectiveness factor is conducted by applying the learned function to complex tissues that were not used as part of the training efforts. Finally, in \S 7, we offer a summary and conclusions.

\section{Microscale Description of Tissue Transport}

At the microscale level of resolution, we define the tissue system as being comprised of 1) an extracellular phase, 2) an intercellular phase, and 3) a cell membrane separating the two phases. 
Although the use of microscale models of tissues using continuum partial differential equations has been promoted for some time \citep{ochoa1987determination, wood1998diffusion}, it has only recently started to be routinely adopted as a method for understanding transport and reactions within complex tissue geometries  \citep{chen2000changes,holter2017,mosharaf2019,ricken2015modeling}.  

\begin{figure*}
\begin{center}
\includegraphics[width=0.8\textwidth]{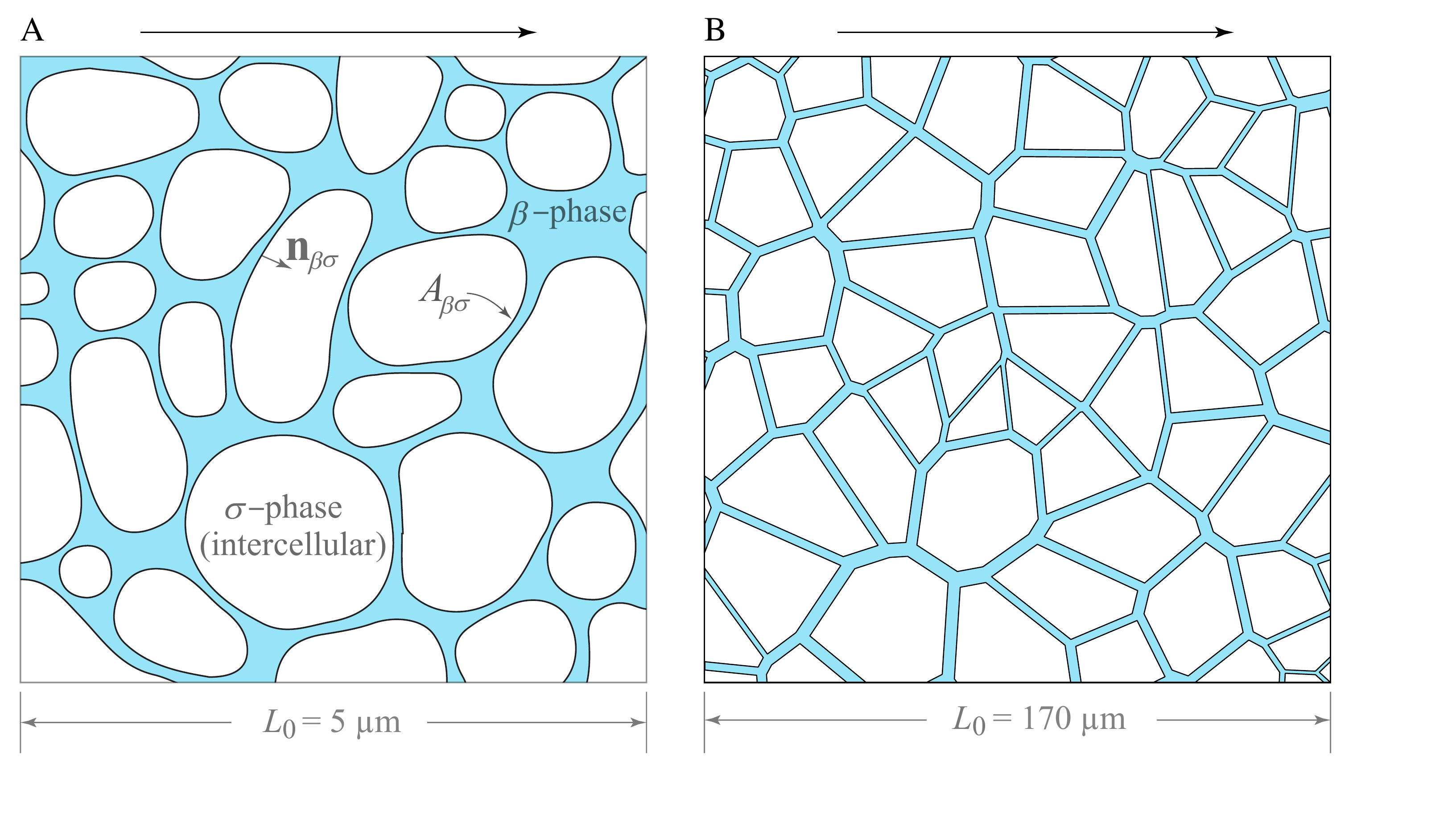}
\caption{Examples of the geometries of the domain $\mathscr{V}$ (with superficial volume $V$) showing extracellular ($V_\beta$) and intercellular ($V_\sigma$) phases, and the cell membrane ($A_{\beta\sigma}$). (A) Image for brain tissue (neuropil), adapted from \citep{chen2000changes}. The extracellular volume fraction is about $\varepsilon_\beta = 0.22$, which is consistent with estimates from \citep{kinney2013extracellular}. (B) Image of engineered liver tissue spheroid, adapted from \citep{leedale2020multiscale}.  Arrows above images show direction of mean flow.}
\label{f:cells}
\end{center}
\end{figure*}

For this research, we have adopted measured 
geometries that are representative of two 
different tissue types (Fig.~\ref{f:cells}): brain cortex tissue \citep{chen2000changes}, and liver lobule tissue \citep{leedale2020multiscale}.  For both tissues,   Michaelis-Menten kinetics 
describe the rate of chemical transformation, which is consistent with a number of previous studies \citep{vendel2019need,leedale2020multiscale}. Similar sub-cell-scale balance equations have been adopted by researchers studying oxygen transport in mesenchymal stem cells \citep{zhao2005effects},  diffusion of chemicals in brain tissue \citep{chen2000changes, vendel2019need, holter2017interstitial}, drug and oxygen transport in liver lobules \citep{leedale2020multiscale}, transport of cytokines in tumors \citep{kim2010interaction}, and movement of solutes in interstitial cell spaces \citep{polacheck2011interstitial} among others.  
 
%

%

To begin the process of upscaling, we first assume that a \emph{representative volume} (RV) of tissue exists.  In short, this means that the tissue structure is organized spatially such that it is sensible to consider macroscopic (or \emph{effective}) representations of the tissue processes.  The details as to the conditions for which an RV exists are usually approached via statistics of the geometric structure.  This issue is discussed in additional detail elsewhere \citep{bachmat1987concept,wood2020primer}.  Here, we assume the REV exists, and
the volume can be divided into the three components, identified as follows (Fig.~\ref{f:cells})  

\begin{equation}
\mathscr{V}({\bf x})=\mathscr{V}_\beta({\bf x}) \cup \mathscr{A}_{\beta\sigma}({\bf x}) \cup \mathscr{V}_\sigma({\bf x})
\end{equation}
where $\mathscr{V}_\beta$ and $\mathscr{V}_\sigma$ represent the extracellular and intracelluar volumes, respectively, and $\mathscr{A}_{\beta\sigma}$ represents the cell membrane separating the two phases.  In the remainder of the paper, a subscript $\beta$ indicates the extracellular phase, whereas a subscript $\sigma$ represents the intracelluar phase. 

At the microscale, for the momentum balance in the extracelluar phase we have adopted the commonly-used Darcy-
Brinkman equations \citep{mosharaf2019} as described by Whitaker \cite[][\S 4.2.6]{whitaker1999} as follows. 

\begin{align}
\intertext{\it Momentum: Extracellular phase}
&& \mu_{eff} \twoform{K}^{-1} {\bf v}_\beta = -\nabla p_\beta+&\rho_\beta {\bf g} +\mu_\beta \nabla^2 {\bf v}_\beta \label{micro8a} \\
&B.C.~1& {\bf v}_\beta &= {\bf 0}, \text{ at cell surface} 
\label{mom}
\end{align}
Here, ${\bf v}_\beta$ is the (intrinsic) fluid velocity of the extracellular phase, $\twoform{K}$ is the Darcy permeability for the extracellular phase, $p_\beta$ is the pressure in the extracellular phase, $\mu_\beta$ is the fluid viscosity, $\mu_{eff}=\phi \mu_\beta$ is the effective viscosity for the Brinkman term, ${\bf g}$ is the gravity vector, and $\phi$ is the fluid volume fraction of within the polymeric substances comprising the extracellular phase.

The microscale mass balance equations can be written as follows

\begin{align}
\intertext{\it Mass: Extracellular phase}
&&    \frac{\partial c_{\beta}}{\partial t} = -{\bf v}_\beta \nabla \cdot c_{\beta} &+\nabla\cdot(\mathscr{D}_{\beta}\nabla c_{\beta})
\label{microe}\\
%
&B.C.~1& -{\bf n}_{\beta\sigma} \cdot \mathscr{D}_{\beta} \nabla c_{\beta} &= -{\bf n}_{\beta\sigma} \cdot \mathscr{D}_{\sigma} \nabla c_{\sigma},\nonumber ~~~\textrm{at cell surface}\\
&B.C.~2& c_{\beta} = & c_{\sigma},~~~\textrm{at cell surface}\\
&I.C.1& c_{\beta}({\bf x},0)& =\mathscr{I}_\beta({\bf x})\\
\intertext{\it Mass: Intracellular phase}
&&\frac{\partial c_{\sigma}}{\partial t} =  \nabla\cdot(\mathscr{D}_{\sigma}\nabla c_{\sigma})& -k_m \frac{c_{\sigma}}{c_{\sigma}+K} \label{nonlin} \\
&I.C.~2& c_{\sigma}({\bf x},0)& =\mathscr{I}_\sigma({\bf x})
\label{microi}
\end{align}
%
Note that external boundary conditions are needed to uniquely complete this description.  In these expressions, $c_{\beta}$ and $c_{\sigma}$ are the concentration of the chemical species of interest in each phase, respectively, $\mathscr{D}_\beta$ and $\mathscr{D}_\sigma$ are diffusion coefficients of the chemical species in each phase, $k_m$ is the maximum velocity kinetic coefficient, and $K$ is the half-saturation constant. 
The presented model is fairly general, and the literature contains data for parameterizing for a number of different physiologically-relevant chemical species with these kinetics.  A sampling of the parameter ranges for this model, available from the literature, is given in Table ~\ref{T:range}. In this table, $r_{\sigma,eff}$ is the effective cell radius.  From this perspective, we do not associate the concentrations $c_\beta$ and $c_\sigma$ with any particular chemical species; rather, we require only that the associated parameters adopted fall within the parameter ranges reported in the literature.

\begin{table}[t]
\centering
    \caption{Data amalgamated from literature sources for brain and liver tissues.  These ranges incorporate both physiologic and experimental conditions. \cite{lettmann2014importance, leedale2020multiscale,sykova2008diffusion,powers2002microfabricated,sharp2019dispersion,debbaut2014,ray2019analysis,maass1997}.}
    \vspace{2mm}
    \begin{tabular}{l|l|l}
        \hline
        ~~Parameter~~                            & ~~~~~~~~~~~~Brain~~~~~~~~~~~~   & Liver  (hepatocyte spheroids) \\
        \hline
        $\mathscr{D}_\beta$ ($\frac{m^2}{s}$)  & $6\times 10^{-10}-20\times 10^{-10}$      & $6\times 10^{-10}-20\times 10^{-10}$ \\ 
        $\mathscr{D}_\sigma/\mathscr{D}_\beta$ & $0.1-1.0$                                 & $0.1-1.0$ \\ 
        $c_{max}$ ($\frac{mol}{m^3}$)          & $0-1.8$                                   & $1\times10^{-3}-1.0$ \\ 
        $\varepsilon_\beta$                    & $0.23-0.49$                               & $0.02-0.41$ \\ 
        $k_m$ ($\frac{mol}{m^3\cdot s^{-1}}$)  & $0.01-1667$                               & $5\times10^{-6}-0.45$ \\ 
        $K$ $\frac{mol}{m^3}$                  & $0.003-528$                               & $5\times10^{-4}-0.14$ \\ 
        $Pe$                                   & $0.01-8.0$                                & $0.01-117$ \\ 
        $r_{\sigma,eff}$ ($m$)                 & $0.47\times10^{-6}$                       & $11.7\times10^{-6}$ \\ 
        $\varphi^2$                            & $0-100$                                   & $0-77$ \\
        $\kappa$ ($m^2$)                       & $2\times 10^{-14}-2\times 10^{-8}$        & $1\times 10^{-10}-7.5\times 10^{-8}$ \\
        \hline
    \end{tabular}
    \label{T:range}
\end{table}


\section{Upscaled Tissue Model}
\label{upscaled}

The macroscale mass balance equations are derived by averaging the microscale equations.  This has been done previously via the method of volume averaging by forming appropriately weighted spatial averages over a representative volume of cells and interfaces (e.g., Fig.~\ref{f:cells}); the results are reported in detail elsewhere \citep{wood1998diffusion,wood2007effective,ochoa1987determination}.  The general result of upscaling in two-phase systems is the development of two macroscale models (one for each phase).  However, such two-phase models are not necessarily the most frequently utilized or the most efficient model to adopt.  In this section we outline the upscaling process and associated constraints where a one-equation model can be derived to represent the tissue mass balance.   In the remainder of this section, we draw on work developed in detail previously by our group \citep{wood1998diffusion, wood2002calculation, wood2007effective, wood2011dispersive, wood2013volume}.  Extensive details regarding the specifics of the upscaling and closure (for generally linearized processes) can be found in those works.

\subsection{One-Equation Model}

While two-equation models can represent a wide range of tissue behaviors, they are significantly less efficient to use in applications than are single-equation models.  One resolution to this problem is to develop a \emph{one-equation macroscale} model that eliminates the coupling between the phases by positing an effective rate of reaction depending only on 1) the model parameters, and 2) the concentration field of the extracellular phase (Eq.~\eqref{effectiveness1}). 
Following results reported previously  \citep{wood1998diffusion,wood2007effective}, the one-equation model takes the form

\begin{equation}
\frac{\partial \langle c_{\beta}\rangle^\beta}{\partial t} = -\langle{\bf v}_\beta\rangle^\beta \cdot \nabla \langle c_{\beta}\rangle^\beta +\nabla\cdot(\twoform{D}^*\cdot \nabla \langle c_{\beta}\rangle^\beta)
+\varepsilon^{-1}_\beta\langle {R} \rangle
\label{one-eq}
\end{equation}
Here, the volume average over the extracelluar phase (the $\beta$ phase occupying $V_\beta$) is defined by the intrinsic average

\begin{equation}
\left.\langle c_\beta \rangle^\beta\right|_{({\bf x},t)} = \varepsilon_\beta^{-1} \int_{{\bf r} \in \mathscr{V}({\bf x})} w({\bf r}-{\bf x})c_\beta({\bf r}, t) I_\beta({\bf r})
dV({\bf r})
\label{volavebeta}
\end{equation}
Note that a \emph{superficial} volume average can also be defined; the two averages are related through the volume fraction by $\langle c_{\beta} \rangle^\beta=\varepsilon_\beta^{-1}\langle c_{\beta} \rangle$. Analogous definitions hold for the intercellular ($\sigma$) phase. 
For the equations above, $\langle{\bf v}_\beta\rangle^\beta$ is the intrinsic average fluid velocity in the extracelluar phase, $\twoform{D}^*$ is the effective dispersion tensor for the extracelluar phase, $\varepsilon_\beta$ is the volume fraction of the extracelluar phase, $\varepsilon_\beta^{-1}\langle {R}\rangle$ is the averaged reaction rate, $w$ is a compact spatial weighting function, and $I_\beta$ is a phase indicator function (where $I_\beta=1$ for spatial points in $V_\beta$, and zero otherwise).   By definition the two volume fractions are related by $\epsilon_\sigma=1-\epsilon_\beta$. 

There are practical reasons that one might adopt the one-equation approach.  For example, in real-world experiments, the extracellular concentration is often more readily measurable than the intercellular concentration (cf. reference \citep[][Chp.~6]{bailey1986} and \citep[][Chp.~10]{truskey2004transport}).  From the computational standpoint, the use of a single-phase model decreases the number of independent balance equations that must be solved (and, hence, the number of degrees of freedom).  Thus, alternatives to the two-equation model have substantial practical value. Macroscale equations using the effectiveness factor approach have been widely adopted for representing the transport and reaction process in tissues.  Examples include the modeling of transport in cartilage \citep{nava2013multiphysics}; substrate transport in mesenchymal cells grown on scaffolds \citep{zhao2005effects}; transport in tumors \citep{dey2018vivo, xie2018modeling}; oxygen transport in bioprinted hepatic sphereoids  \citep{khakpour2017oxygen}, oxygen transport in muscle tissue \citep{dasika2011reaction}, and substrate transport in hollow-fiber supported tissues \citep{shipley2012fluid, chapman2017mathematical}.

\subsection{Computing the Effectiveness Factor and Effective Dispersion Tensor}

The superficial average reaction rate within the volume, $\mathscr{V}({\bf x})$, is found by computing the integral

\begin{align}
     \langle {R}\rangle &= 
    -k_m \int_{{\bf r} \in \mathscr{V}({\bf x})} w({\bf r}-{\bf x})I_\sigma({\bf r}) \frac{c_\sigma({\bf r})}{c_\sigma({\bf r}) +K} \,\, dV({\bf r})
    \label{micro_effective_R}
\end{align}
This provides a scheme for defining the effectiveness factor. We begin with the \emph{proxy} average reaction rate expression, $R_0$

\begin{align}
 R_0 =-\left({\varepsilon_\sigma} k_m \frac{\langle c_{\beta}\rangle^\beta}{\langle c_{\beta}\rangle^\beta+ K} \right)
\label{R_0}
\end{align}
Clearly, $R_0$ is generally not the correct reaction rate (although it may be correct in certain limiting regimes).  This definition is motivated by simply replacing the concentration terms (inside the intracellular phase, $c_\sigma$) appearing in the non-linear reaction term in Eq.~\eqref{nonlin} with the intrinsic average concentration, $\langle c_{\beta}\rangle^\beta$, for one-equation model. It is apparent that this definition needs a correction term that can generate the actual reaction rate given by Eq.~\eqref{micro_effective_R}. Hence, the expression is adjusted by the effectiveness factor, $\eta$.  This defines the effective rate of reaction by the relationship \citep{shuler1972diffusive}
\begin{align}
\langle{R}\rangle &= \eta~R_{0} 
\label{effectiveness_factor} \\
&= -{\eta}\left({\varepsilon_\sigma} k_m \frac{\langle c_{\beta}\rangle^\beta}{\langle c_{\beta}\rangle^\beta+ K} \right)
\label{effective_def}
\end{align}

In this work, there are two effective parameters to be determined via closure, $\twoform{D}^*$ and $\eta$, using the solutions of Eqs.~\eqref{mom}-\eqref{microi}.  The closure for the effective dispersion coefficient, $\twoform{D}^*$, is a linear problem that has been extensively studied \citep{whitaker1999}.  For isotropic media, the dispersion tensor is primarily a function of a two parameters, the P\'eclet number and the extracellular volume fraction, $\epsilon_\beta$.   The computation for $\twoform{D}^*_A$ can be accomplished following a conventional analysis as outlined by Whitaker \citep[][Chp.~3]{whitaker1999}. For completeness, this computation is outlined in the Appendix.

The challenge at this juncture is to determine an appropriate method for predicting the effectiveness factor. For linear (or linearized) problems, closure can be achieved by a sequence of algebraic manipulations of the microscale and macroscale balance equations (cf. reference \citep{wood2007effective}). For nonlinear problems, there are no general methods; closures are usually developed through various linearizations of the problem to provide asymptotically valid solutions. If the goal of the closure is to obtain results that are valid under general conditions, one must resort to numerical approaches to compute $\eta$.  

Unlike the effective dispersion tensor, the effectiveness factor is a complex function of the average concentration fields, the kinetic rate parameters ($k_m$ and $K$) and the transport parameters $\twoform{D}^*_A$, the average velocity $\langle{\bf v}_\beta\rangle^\beta$, and the geometry of the problem.  Thus, the function defining $\eta$ is of significantly higher dimension (i.e., it is described by more than one or two independent variables) than that defining the effective dispersion tensor.  Parsing out the particular independent variables on which $\eta$ depends is discussed under the section on feature engineering; the details of feature engineering are described in \S\ref{learning}.

\subsection{Algorithm for Computing the Effectiveness Factor}
\label{effectiveness}

For a specified realization of the tissues involved in this work, the effectiveness factor can be numerically computed by solving the microscale balance equations (Eqs.~\eqref{mom}-\eqref{microi}). 
To compute $\eta$, one needs to first compute the microscale concentration fields over a representative region, $\mathscr{V}({\bf x})$, for a specified set of parameters. One algorithm for computing $\eta$ is as follows.

\begin{enumerate}
    \item Within an averaging volume, compute the actual rate of reaction $\langle {R} \rangle$, from the microscale concentration field, as indicated by Eq.~\eqref{micro_effective_R}.
    \item Compute the intrinsic average concentration, $\langle c_\beta\rangle^\beta$, in the extracellular phase within $\mathscr{V}({\bf x})$.
    \item Compute $R_{0}$ using Eq.~\eqref{R_0}.
    \item Compute the effectiveness factor, from $\eta = {\langle{R}\rangle}/{R_{0}}$.
\end{enumerate}

There is one additional facet to this computation that requires brief discussion.  In general, the effectiveness factor is a function of the (generally transient) concentration field. However, there are often significant differences in the magnitude of the characteristic time scales for the microscale ($t^*$) and macroscale ($T^*$) processes (cf. references \citep{whitaker1999,wood1998diffusion, wood2007effective}) such that $t^* \ll T^*$.  In other words, a small perturbation in any macroscale parameter (on time scale $T^*$) is rapidly relaxed at the microscale (taking time $t^*$).  In such situations, the microscale problem can be treated as a quasi-steady one \cite{wood2007effective}.  This indicates that the effectiveness factor can be computed from the steady-state versions of Eqs.~\eqref{microe}-\eqref{microi}.  Regardless of this generality, in many applications the processes involved do occur at steady state.  In the remainder of this work, we will focus on predicting the effectiveness factor for steady-state conditions.

\section{Physics-Driven Feature Engineering and Generation of Example Data Ensemble}
\label{learning}

The ML problem for this application is a supervised learning process. The the first step in the process is to establish a vector of independent variables, $\boldsymbol{x}$, that determine the unknown function of the effectiveness factor, i.e., $\eta = f^*(\boldsymbol{x})$ (N.B. the distinction between the feature vector, $\boldsymbol{x}$ and the coordinate vector, ${\bf x}$).  These independent variables are known as \emph{features}; we assume that there are $k$ such features that are identified.  The array of all $j$ examples of $k$ features is denoted by ${\boldsymbol{X}}=x^{(j,k)}$.  

For this work, the features can be classified into two main categories: 1) features that are independently known parameters (or parameter groupings) from the nondimensionalized microscale balances and boundary conditions describing the system; we refer to this class of parameters as \emph{explicit} physics driven features, and 2) features that are macroscale parameters, macroscale concentrations, or macroscale concentration gradients that can be computed (by averaging) from the training set of microscale solutions; we refer to this second class as \emph{implicit} physics driven features. These two types of the features are strictly driven by the physics of the problem.  Note that the features should not depend explicitly upon the microscale concentration fields since this is the dependent variable we are attempting to eliminate via closure.  Also, note that each additional feature adds one additional constraint to the loss function that is being minimized during the training process. Ideally, then, the addition of useful features results in a decrease in the error norm.

As a matter of notation, we define the training set of features and target values by $\boldsymbol{\eta}=f(\boldsymbol{X};\boldsymbol{\theta})$, where $\boldsymbol{\theta}$ is an array of parameters (weights and biases) used in the neural network (described in the next section).  Note that $f^*$ is the (unknown) function that is to be estimated, whereas $f$ represents the observed value from the training data set; in the material following, we will use $\hat{f}$ to indicate the estimates of $f^*$ from the MLP network.

\subsection{Explicit parameters}

Nondimensionalization of the microscale balance equations leads to a learning framework that is independent of the physical dimensions of the problem; this allows us to learn the model on a geometry from which the \emph{scale} has been removed, increasing the generalizability of the learned model. The choice of brain and liver geometries was based, in part, on the large difference in their scale. If a model trained on nondimensionalized data is able to describe the effectiveness factor for both systems, then it suggests that the rescaling has captured, to some extent, universal behavior. 
The microscale balance equations were nondimensionalized as follows.

\vspace{-2mm}
\begin{align}
&&    \frac{\partial C_{\beta}}{\partial \tau} &= -Pe\frac{{\bf v}_\beta}{U} \nabla \cdot C_{\beta} +\nabla^2 C_{\beta}\\
&I.C.1& C_{\beta}({\bf Z},0)& ={I}_\beta({\bf Z})
\end{align}
\begin{align}
&B.C.~1& -{\bf n}_{\beta\sigma} \cdot  \nabla C_{\beta} &= -{\bf n}_{\beta\sigma}\cdot (D_r   \nabla C_{\sigma}),~\textrm{at cell surface} \\
&B.C.~2& C_{\beta} &= C_{\sigma},~\textrm{at cell surface}
\end{align}
\begin{align}
&&    \frac{\partial C_{\sigma}}{\partial \tau} &=  D_r\nabla^2 C_{\sigma} -\varphi^2  \frac{C_{\sigma}}{C_{\sigma}+1}     \\
&I.C.2& C_{\sigma}({\bf Z},0)& ={I}_\sigma({\bf Z})
\end{align}
where here
\begin{align*}
    C_{\beta} &= \frac{c_{\beta}}{K};&\quad
    C_{\sigma} &= \frac{ c_{\sigma}}{ K};&\quad
    D_r &=\frac{\mathscr{D}_\sigma}{\mathscr{D}_\beta};&\quad
    Pe &= \frac{U r_{\sigma,eff}}{\mathscr{D}_{\beta}} \\
    \varphi^2 &= \frac{ k_m r^2_{\sigma,eff}}{ K \mathscr{D}_{\beta}};&\quad
    \tau &= \frac{r^2_{\sigma,eff}}{\mathscr{D}_{\beta} t};&\quad
    {\bf Z} &= \frac{\bf z}{r_{\sigma,eff}};&\quad&~
\end{align*}
Here we adopt the coordinate system ${\bf z}=(x,y,z)$ (and its non-dimensional form ${\bf Z}=(X,Y,Z)=(x/r_{\sigma,eff},y/r_{\sigma,eff},z/r_{\sigma,eff})$), where the $z-$axis is aligned with the mean direction of flow.  Note that here the gradient operator is the nondimenaional one $\nabla \equiv \frac{\partial}{\partial {\bf Z}}$. By the symmetry of the problem, we have that $\nabla\langle C_\beta\rangle=(0,0,\partial \langle C_\beta\rangle/\partial Z)$.  Finally, the value of $U$ is the intrinsic volume average of the velocity over the entire domain,
i.e., $U = (\langle {\bf v_\beta} \rangle^\beta \cdot \langle {\bf v_\beta} \rangle^\beta)^{\tfrac{1}{2}}$.

The effects of reaction rate, convection, and diffusion forces are manifest in the dimensionless numbers $\varphi^2 = { k_m r^2_{\sigma,eff}}/{(K\mathscr{D}_{\beta})}$ (computed in the cell phase), $Pe$ (computed in the extracellular phase), and $D_{r}=\mathscr{D}_{\sigma}/\mathscr{D}_{\beta}$, respectively; these quantities are natural choices for the feature set.  

\subsection{Implicit parameters}

The additional selection of the \emph{average concentration} and \emph{gradient of the average concentration} as features can be motivated in two ways.  First, empirically we know that the effectiveness factor depends directly on the average concentration, as specified by Eq.~\eqref{effectiveness_factor}.  One might, on this basis, be motivated to try using the first few terms of a Taylor series expansion as features (equivalent to using $\langle C_\beta\rangle$ and $\partial \langle C_\beta\rangle/\partial Z$ as features).  More directly, however, it can be shown that the average concentration and its gradient appear in the Greens function solution for the effective parameters in linearized versions of this problem.  For example the quasi-steady closure for the linearized version of this problem (found by subtracting the average equation from the microscale equation) and the internal boundary condition can be expressed as \citep{wood2007effective}

\begin{align}
 &&   {\bf v}_\beta \cdot \nabla \tilde{c}_\beta 
 + \underbrace{\tilde{\bf v}_\beta \cdot \nabla \langle c_\beta \rangle^\beta}_{\text{source}}
 &= \nabla \cdot(\mathscr{D}_\beta \nabla \tilde{c}_\beta)
 +\underbrace{\varepsilon_\beta^{-1} a_v \frac{k_m}{K}\langle c_\beta \rangle^\beta}_{\text{source}} +\varepsilon_\beta^{-1} a_v \frac{k_m}{K}\frac{1}{A_{\beta\sigma}}\int_{{\bf r} \in \mathscr{A}_{\beta\sigma}}\tilde{c}_\beta \,dA \\
&B.C.~1 & -\mathscr{D}_\beta \nabla \tilde{c}_\beta \cdot {\bf n}_{\beta\sigma} -\frac{k_m}{K}\tilde{c}_\beta 
&= \underbrace{\mathscr{D}_\beta \nabla \langle c_\beta\rangle^\beta \cdot {\bf n}_{\beta\sigma}}_{\text{source}}
+\underbrace{\frac{k_m}{K}\langle c_\beta\rangle^\beta}_{\text{source}}
\end{align}
The terms denoted by \emph{source} in this expression are macroscale terms that arise in the microscale closure problem.  When a integral solution in terms of Green's functions is constructed, these source terms appear in convolution integrals with the Green's functions (see \citep{wood2013volume} for additional details).  Ultimately, the behavior of the effectiveness factor is determined by the solution to this closure problem. More importantly, including the $\langle c_\beta \rangle^\beta$ to the feature set preserves the mass balance in the system. These establish the rational for including both $\langle c_\beta \rangle^\beta$ and $\nabla \langle c_\beta \rangle^\beta$ as features for the prediction of $\eta$.


\subsection{Representative geometry}

The status of the system geometry as a feature is a somewhat complex problem.  In many important processes in biological systems (e.g., flow in vascular networks), one would expect the details of the geometry to potentially be an important component of the feature set describing the behavior of the system.  For convection-diffusion problems in nearly isotropic (e.g., close to circular/spherical) geometries, it has been observed that the volume fraction alone provides a good representation of the geometrical information.  While this is not to say that additional geometrical details could not be extracted by incorporating additional learning methods (e.g., using convolutional neural networks to extract geometrical information from images of representative volumes e.g., (see \citep{li2020reaction, wu2019predicting}), for the geometries expressed by these tissues, such additional efforts are an active area of research.

Because the geometry is reflected primarily by the porosity for this system, we adopted a \emph{representative} geometry to model the system.  The representative geometry provides sufficient similarity to the actual geometrical structure such that it captures the essential features of the more realistic geometries being modeled (such as those presented in Fig.~\ref{f:cells}), but is simple enough such that the solutions to the unit cell problems are not overly costly to solve numerically.

\begin{figure}[t]
\begin{center}
\includegraphics[width=0.6\textwidth]{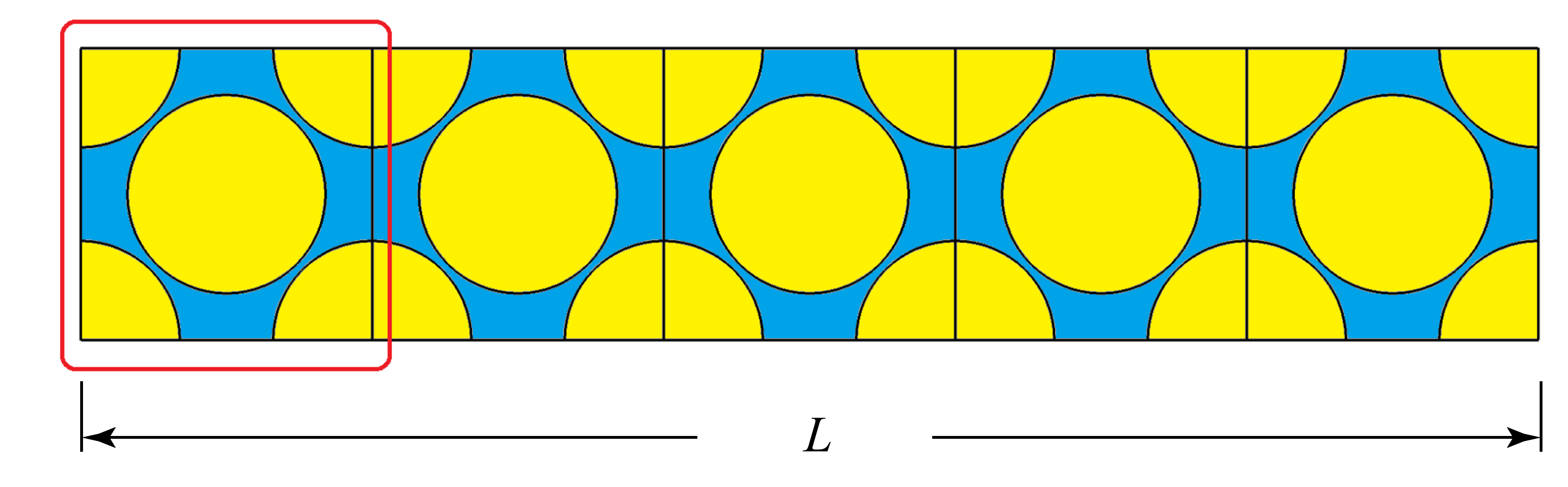}
\caption{The until cell depicted inside the red box and an array of the unit cell. The flow and cell phases are distinguishable by color. The square's length is $L_0=3\times 10^{-6} (m)$ and the circle's radius is $r_{\sigma,eff}=1.02 \times 10^{-6} (m)$.}
\label{f:RepGeo}
\end{center}
\end{figure}

As with other elements of the learning process, the validity of any particular representative structure as a proxy for the actual geometry of the tissues involved can only be checked heuristically.  The representative geometry adopted for this work is the simple set of unit cells illustrated in Fig.~\ref{f:RepGeo}.  While these structures are quite simple, our primary focus is to correctly capture the intercellular diffusion and reaction process.  It has been illustrated a number of times that isotropic diffusion is well represented by a periodic array of circles (or spheres in 3D), and is stronger a function of the volume fraction than the particular geometry employed \citep{quintard1993}; similarly, simple unit cells have been used previously to compute the effectiveness factor \cite{wood2007effective}.   To avoid the strong influence of the Dirichlet boundary conditions imposed at the inlet of the system, we used a sequence of five unit cells, Fig.~\ref{f:RepGeo}, for each simulation, while the data from the first cell was discarded.  Note that the values for $\langle C_\beta \rangle^\beta$ were determined using Eq.~\eqref{volavebeta}, where the weighting function, $w$, was taken to be a uniform (top-hat) function with width equal to $2\,r_{\sigma,eff}$.  Estimates for the derivative $\frac{\partial}{\partial Z} \langle C_{\beta}\rangle^\beta$ were computed by a centered finite difference after computing the average concentration $\langle C_{\beta}\rangle^\beta$ in each cell.

\subsection{Computability and consistency requirements}

Regardless of any other arguments supporting the set of features described above, it is necessary that the set of features be at least consistent with the available information and physics of the system being described.  The explicit parameter set given by $\varphi^2$, $Pe$, $D_r$ clearly meet this requirement.  The representative geometry proposed for this problem has significant support from previous work done in volume averaging \citep{whitaker1999, wood1998diffusion, wood2002calculation} for convection-diffusion-reaction in nearly isotropic systems.   This, provides strong motivation for the use of $\varepsilon_\beta$ as a feature variable. 

The two macroscale variables (the average concentration and gradient of the average concentration) have clear motivation for use as feature variables when examining the source terms in the differential balances that define the effectiveness factor.  
From a more practical perspective, one can easily compute estimates for the averaged concentration $\langle C_{\beta}\rangle^\beta$ and average gradient, $\frac{\partial}{\partial Z} \langle C_{\beta}\rangle^\beta$ directly from the ensemble of microscale simulations used to generate the target data.  In Fig.~\ref{f:RepGeo} we illustrate one example of a unit cell array used in the ensemble of microscale computations. The average concentration is estimated cell-wise by conducting the appropriate integrations over each cell in the array.  For determining estimates of the gradient, centered finite differences of the averages can then be employed on the the spatially averaged concentrations defined for the array. 
Based on these considerations, the following feature set was adopted for representation in the MLP: 1) $\varepsilon_\beta$, 2) $\varphi^2$, 3) $Pe$, 4) $D_r$, 5) $\langle C_{\beta}\rangle^\beta$, and 6) $\frac{\partial}{\partial Z} \langle C_{\beta}\rangle^\beta$.

We note that each of the chosen features has a high correlation with $\eta$ (Fig.~\ref{f:correction_matrix}), indicating that each contributes significantly to the reduction in variance for the model.  Because the features are largely bounded by physically relevant parameters, no specific feature selection algorithms (e.g., LASSO regression \citep{goodfellow2016deep} for dimension reduction) were required.

\subsection{Generation of the example data ensemble} 

There are two common challenges in generating the training data such that the training process properly reflects the underlying functional relationship.

\begin{enumerate}
    \item The distribution of the dataset should not be highly unbalanced.  When the distribution of the dataset is biased or skewed, the results can be poor model performance because important subdomains of the feature space have relatively small representation in the data set.
    \item The dataset should sample the feature space densely enough so that the neural network predicts the value of $\eta$ with high fidelity for any set of given points in the feature space.
\end{enumerate} 

The first of these issues can be handled by careful generation of the feature space samples.  For example, features are usually normalized to occur on an interval from zero to one \citep{friedman2001elements}.   For nonlinear problems, the distribution of the feature space must be determined empirically; however, striving for close-to-uniform distributions for each feature can provide a reasonable starting point. For the second issue, there is a trade off between the sampling density in the ensemble of examples, and the potential for overfitting the data.  Generally, more densely sampling the feature space will allow the learned model to have good fidelity (low error) while avoiding overfitting \citep{chollet2018deep}.  

Table~\ref{T:range} summarizes the range of physiological and experimental data found in the literature; these establish reasonable domains for each feature. Based on this table, we set the domain for training as follows: $0.25 < \varepsilon_\beta < 0.85$, $0.01 < \varphi^2 < 100$, $0.01 < Pe < 100$, $0.01 < D_r < 1$, and $ 0.1 < C_{max} < 10$ to ensure that they cover the reported data for brain and liver. We use the MATLAB's build-in \emph{Latin hypercube} sampling function to generate random distributions, as close to uniform as practical, for $\varepsilon_\beta$, $D_r$, and $C_{max}$.  

In order to get a close-to-uniform distribution for the target value, $\eta$, it was necessary to adopt \emph{non-uniform distributions} for $\varphi^2$ and $Pe$.  This was achieved heuristically by trying various non-uniform distributions for $\varphi^2$ and $Pe$ to find a distribution that returned a close-to-uniform distribution for $\eta$. 

To conduct the learning process, we need an appropriate ensemble of known values of $\eta$ that are associated with a known vector of feature values.  To generate such examples, we computed the microscale solutions to the balance equations over the representative geometry illustrated in (Fig.~\ref{f:RepGeo}).  Realizations of feature sets were generated randomly from the distributions reported in the previous section.  For each such sample of the feature space, we employed the finite element software COMSOL Multiphysics 5.5\textsuperscript{\textregistered} to solve the steady-state balance equations implemented on the representative geometry, and compute the corresponding target value, $\eta$. 

For the flow problem, the external boundary conditions were periodic on the surfaces perpendicular to mean flow, and specified pressures at the inlet and outlet surfaces.  For the mass transport and reaction problem, periodic conditions were used on the external surfaces perpendicular to mean flow, the inlet boundary was set as a specified concentration, and the outlet boundary was specified by a zero-concentration-gradient condition.  Internal boundaries were as specified in Eqs.~\eqref{microe}-\eqref{microi}. 
We performed a convergence analysis based on Richardson extrapolation on the simulations with the highest convection as well as lowest and highest reaction rates following Roache \citep{roache1994perspective} in order to ensure that the numerical results are stable. We computed the grid convergence index (GCI), which provides a bound on the estimated error of the numerically converged solution, for the simulations. We imposed the condition that the GCI be on the order of $1\times 10^{-4}$ or less, indicating a grid-independent solution.  For each simulation (each with a unique set of feature data) we computed the effectiveness factor as described in \S \ref{upscaled}.\ref{effectiveness}.

\begin{figure}
\begin{center}
\includegraphics[width=0.5\textwidth]{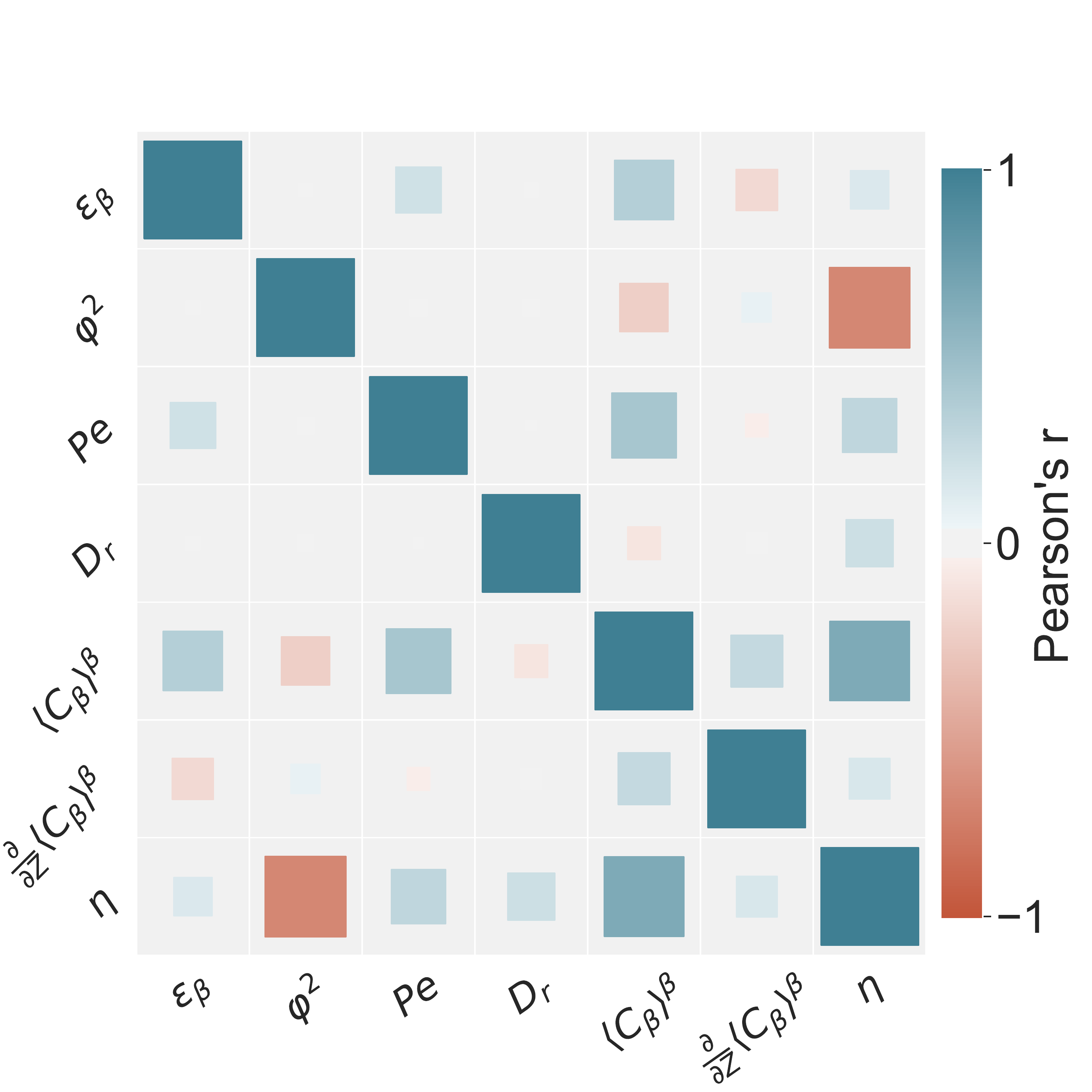}
\caption{The Pearson's correlation between each pair of features and target set. Note that the target space $(\eta)$ is fully covered.}
\label{f:correction_matrix}
\end{center}
\end{figure}

To asses the quality of our selected feature set, we computed correlations among the features for the entire set of simulations comprising the training data set.  In Fig.~\ref{f:correction_matrix} the Pearson's correlation between each pair of features is illustrated. Given a pair of features, $X^k=x^{(j,k)}$ and $Y^\ell=x^{(j,\ell)}$, the bivariance correlation, also know as the Pearson product-moment correlation coefficient, is defined as
\begin{align}
    \rho(X^k,Y^\ell) = \frac{\text{cov}(X^k,Y^\ell)}{\sigma_{X^k} \sigma_{Y^\ell}}
\end{align}
where $\text{cov}$ is the covariance and $\sigma^k$ is the standard deviation for feature $k$. The positive and negative linear correlation are represented as blue and red squares; the bigger square, the strongest linear correlation. From the last row (or last column), it is evident that 1) all the selected features substantially influence the target value; the corresponding learned weights for each feature will not be close to zero throughout the learning process; and 2) the order in which the features affect $\eta$ is $ \varepsilon_\beta < \frac{\partial}{\partial z} \langle C_{\beta}\rangle^\beta < D_r <  Pe < \langle C_{\beta}\rangle^\beta < \varphi^2 $ for the generated dataset. These relations show the significance of the source term, $\langle C_{\beta}\rangle^\beta$, in feature set as it preserves the mass balance.

\section{Feed Forward Neural Network}
\begin{figure}[t]
\begin{center}
\includegraphics[width=0.6\textwidth]{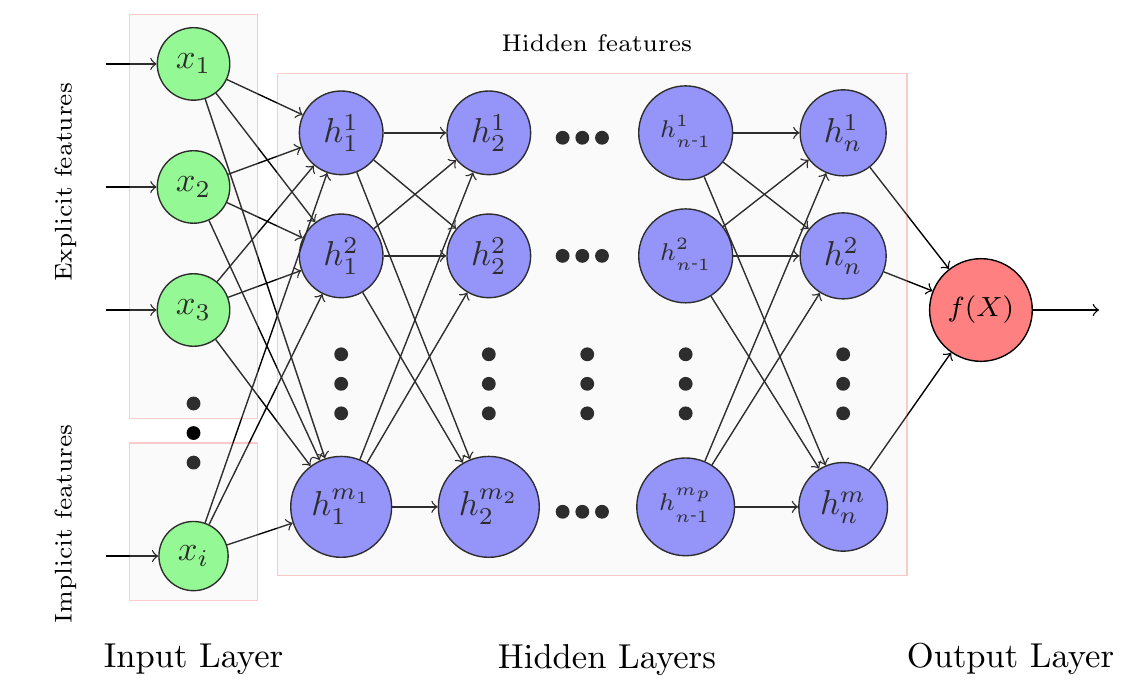}
\caption{A schematic of a multilayer perceptron designed for regression task with $n$ hidden layers each containing a different number of nodes (neurons) denoted by $m$.  The MLP used in this work contained 5 hidden layers, each with a nonlinear ReLu (rectified linear unit) transformation.  The number of neurons in each of the hidden layers was equal to 1024, 512, 256, 64, and 16, respectively.}
\label{f:MLP}
\end{center}
\end{figure}

The MLP is instantiated by a plain stack of layers, knows as a sequential model, where each layer has exactly one input and one output tensor. Our designed network consisted of an input layer with a uniform kernel initializer, five dense fully connected hidden layers, and an output layer. The number of neurons in each layer were 1024, 512, 256, 64, 16, respectively. A graphical representation of the network is given in Fig.~\ref{f:MLP}. For each hidden layer a rectified linear unit (ReLU) function was utilized as activation function to nonlinearlly transform the input functions. The input data were normalized to ensure that they are contained in the same order of magnitude to help prevent overfitting \citep{goodfellow2016deep}. An adaptive moments (``Adam'') gradient-based optimization scheme was adopted, with a learning rate (multiplier, $\epsilon$, in the gradient-based optimizer) equal to $\epsilon=10^{-3}$ and a time-based decay rate of learning rate/iteration number. For $N$ training examples, the mean absolute percentage error ($MAPE$), defined as

\begin{equation}
  MAPE= \frac{1}{N} \sum_{j=1}^{j=N}\left|\frac{\eta^{(j)}-{f}(\boldsymbol{x}^{(j)},\boldsymbol{\theta})}{\eta^{(j)} }\right|\times 100
\end{equation} 
(where $\eta^{(j)}$ is the true value from the data vector, and ${f}(\boldsymbol{x}^{(j)};\boldsymbol{\theta})$ is the value predicted for input ${\boldsymbol{x}^{(j)}}$ from the network), was used as the \emph{loss function}. Note that the model seeks to minimize the defined loss function during the training process. 

For training the network, of the total data set, 76\% was used as training data, 4\% as validation, and 20\% as test data; the later was used to determine the generalization error \citep{goodfellow2016deep} based on the mean squared error (MSE) defined by

\begin{equation}
  MSE= \frac{1}{N} \sum_{j=1}^{j=N} {(\eta^{(j)}-{f}(\boldsymbol{x}^{(j)};\boldsymbol{\theta}))^2}
\end{equation}
Note that neither the validation nor the test dataset contribute to the learning. Since the dataset is randomly split into the training, validation, and test components with each run, it yielded very slightly different results; as a result, we present the best results obtained from 10 runs. The tuning of the network's hyperparameters (number of layers, type of activation function, dimensionality of the output space for each layer, etc.) was achieved during the validation step, primarily via heuristics.  
The average running time for 2000 epochs was about 1200 (s) on a Geforce GTX 1080 Ti GPU. We used Python 3.7 and Tensorflow 2.3.0 for developing the architecture.

\begin{figure}
\begin{center}
\includegraphics[width=0.45\textwidth]{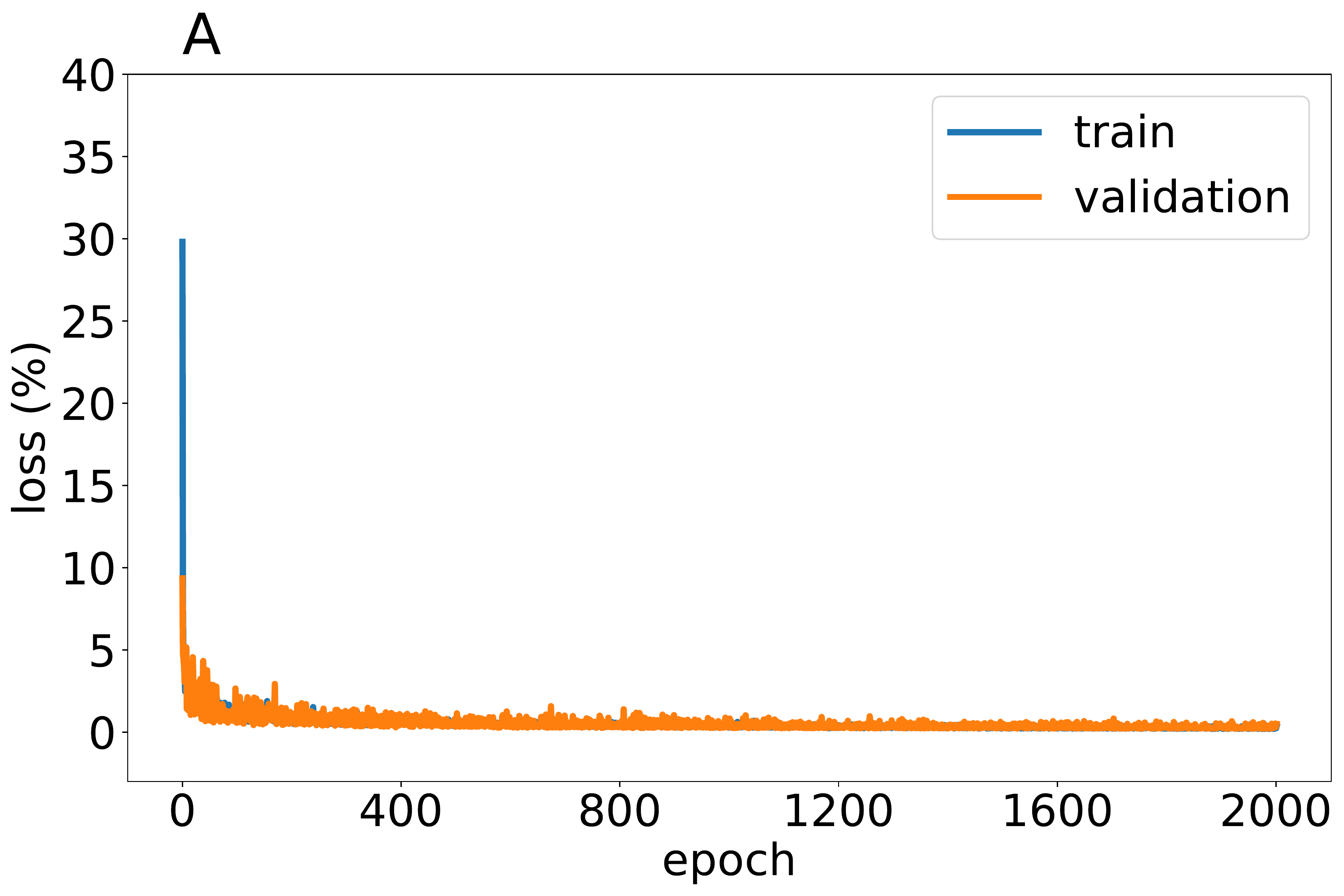}
\hspace{2mm}
\includegraphics[width=0.45\textwidth]{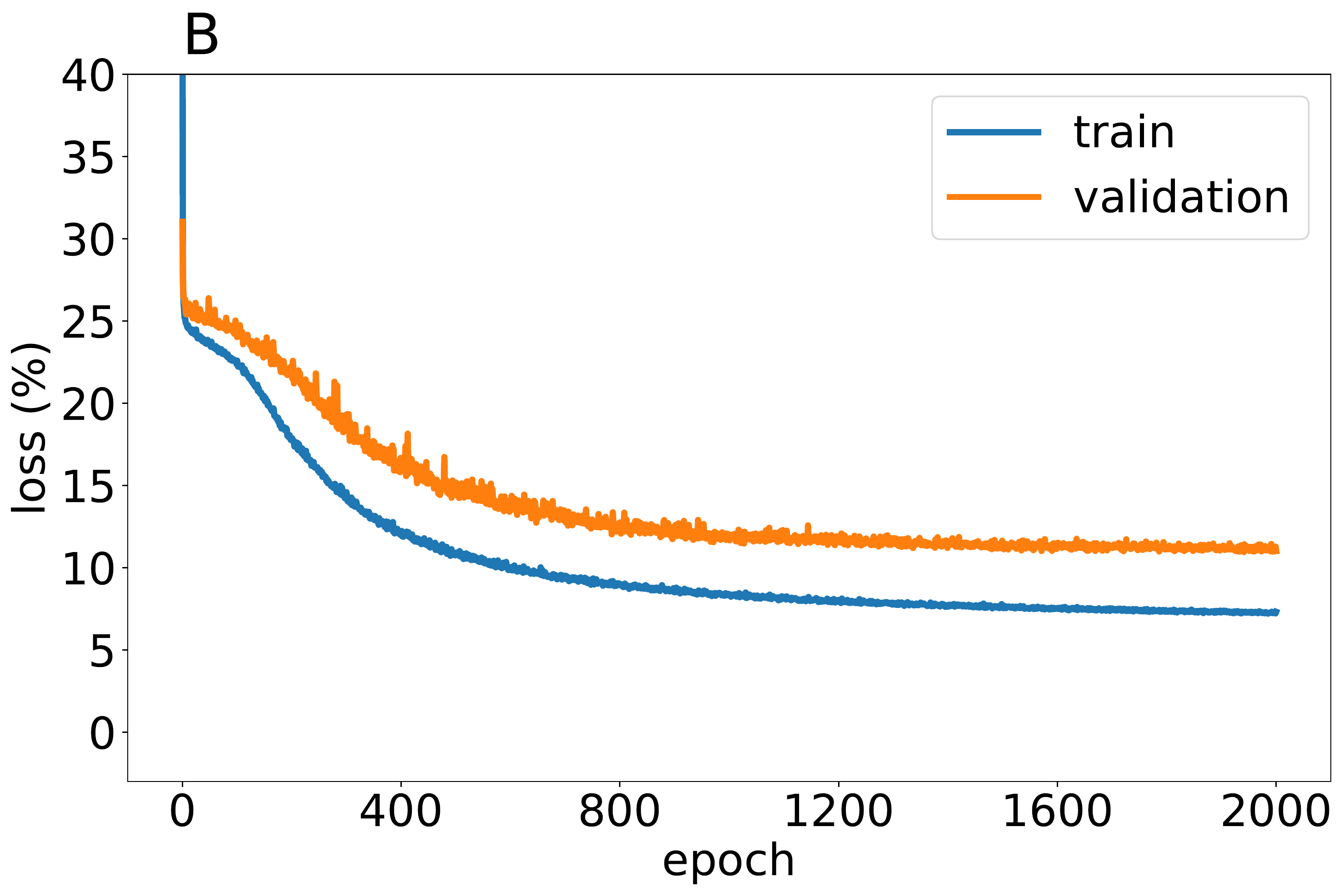}\\
\includegraphics[width=0.45\textwidth]{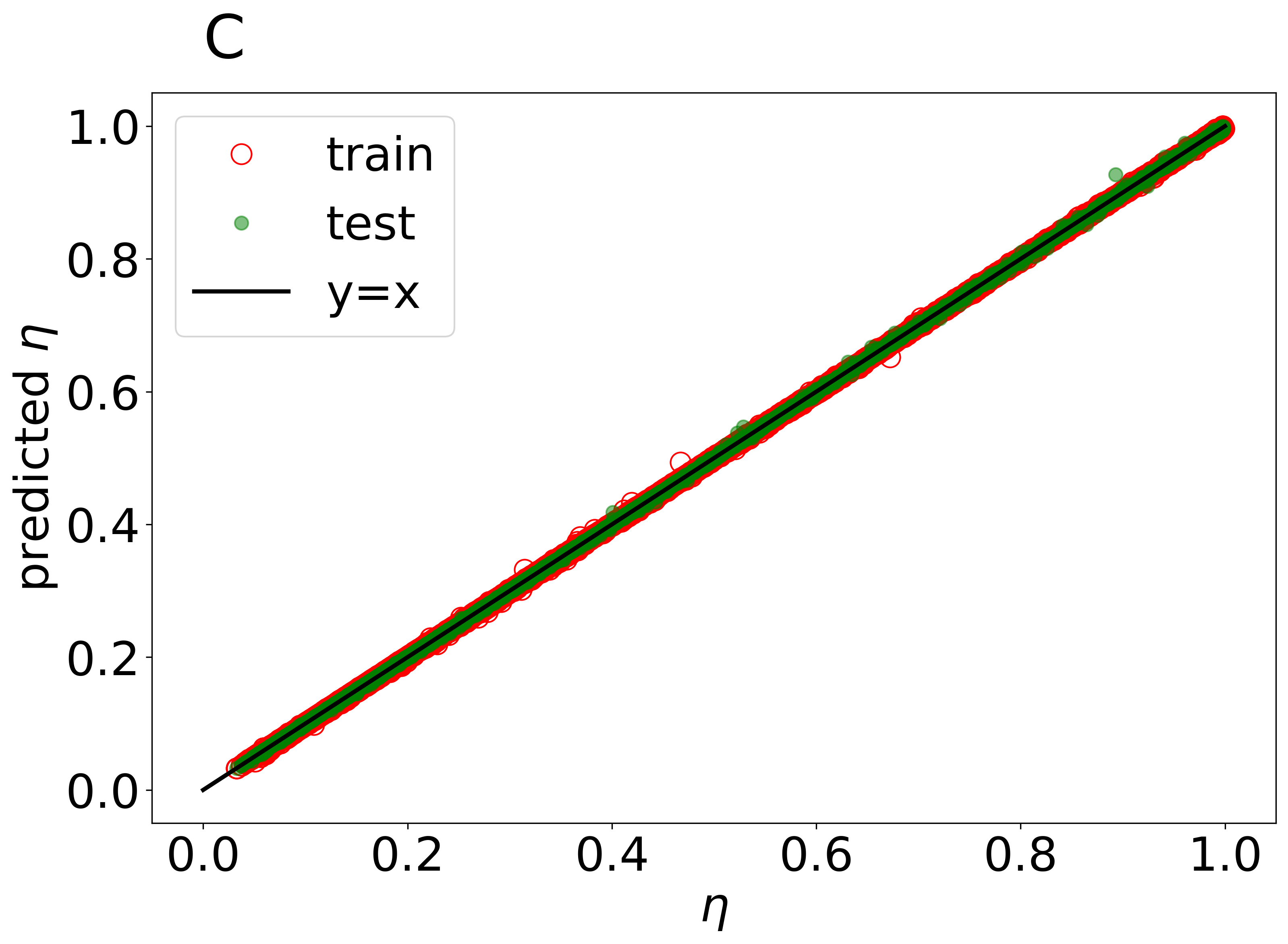}
\hspace{2mm}
\includegraphics[width=0.45\textwidth]{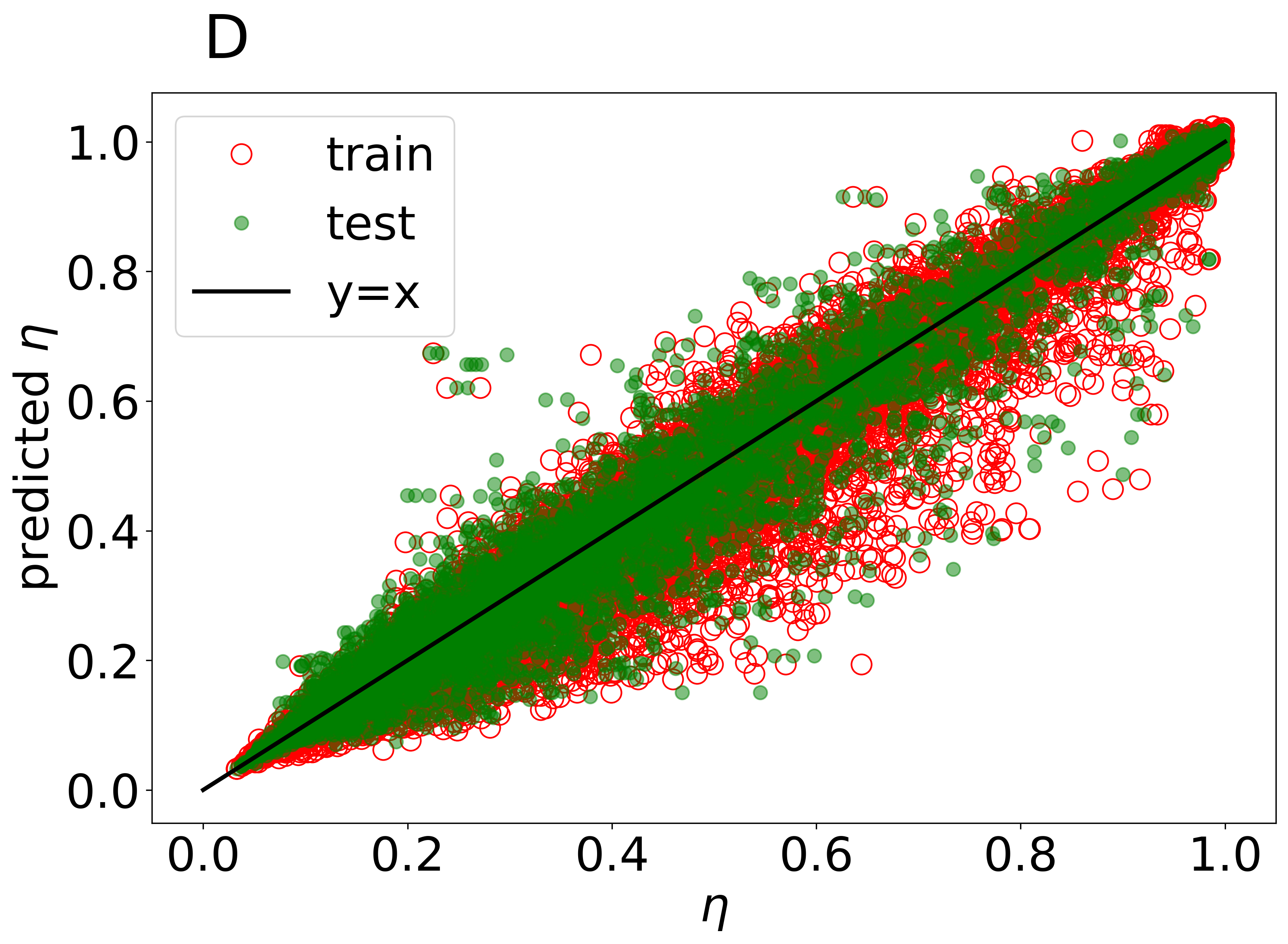}

\caption{(A) The MAPE loss function for the case where both explicit and implicit features included. (B) The MAPE loss function for the case where only explicit features included.  (C) The predicted $\eta$ by MLP (at 2000 epochs) vs. ground truth $\eta$, determined by microscale simulation, for the case where both explicit and implicit features included.  (D) The predicted $\eta$ by MLP (at 2000 epochs) vs. ground truth $\eta$, determined by microscale simulation, for the case where only explicit features included.}
\label{f:loss}
\end{center}
\end{figure}

In Fig.~\ref{f:loss}(A-B), the loss function and MSE versus the number of epochs is illustrated.  The primary significance of these plots is to show that introducing the implicit features can drastically decrease the error and, hence, improve the network prediction results (left-hand-side figures compared to right-hand-side ones). 

The use of implicit features (the source terms in the closure for this problem) is a somewhat novel aspect of this work. The significance of the source terms in unveiling the underlying physics of the complex transport phenomena has been specifically addressed \citep{taghizadeh2020preasymptotic}. We examined the influence of including the implicit features $\langle C_{\beta}\rangle^\beta$ and $\partial \langle C_{\beta}\rangle^\beta/\partial Z$ on the resulting error in the trained network; These results are plotted in Fig.~\ref{f:loss}(B). It is clear from these results that the inclusion of the physically-motivated source terms dramatically improves the performance of the trained network. Without the implicit features, the MAPE and MSE error are 11\% and 4.0$\times 10^{-3}$, respectively.  When these features are included, the MAPE 0.55\% and MSE 7.1$\times 10^{-6}$.  This implies that both classes of features (explicit and implicit) are useful for obtaining a model with low training error.




\subsection{Training Data Set Size}

\begin{figure}
\begin{center}
\includegraphics[width=0.45\textwidth]{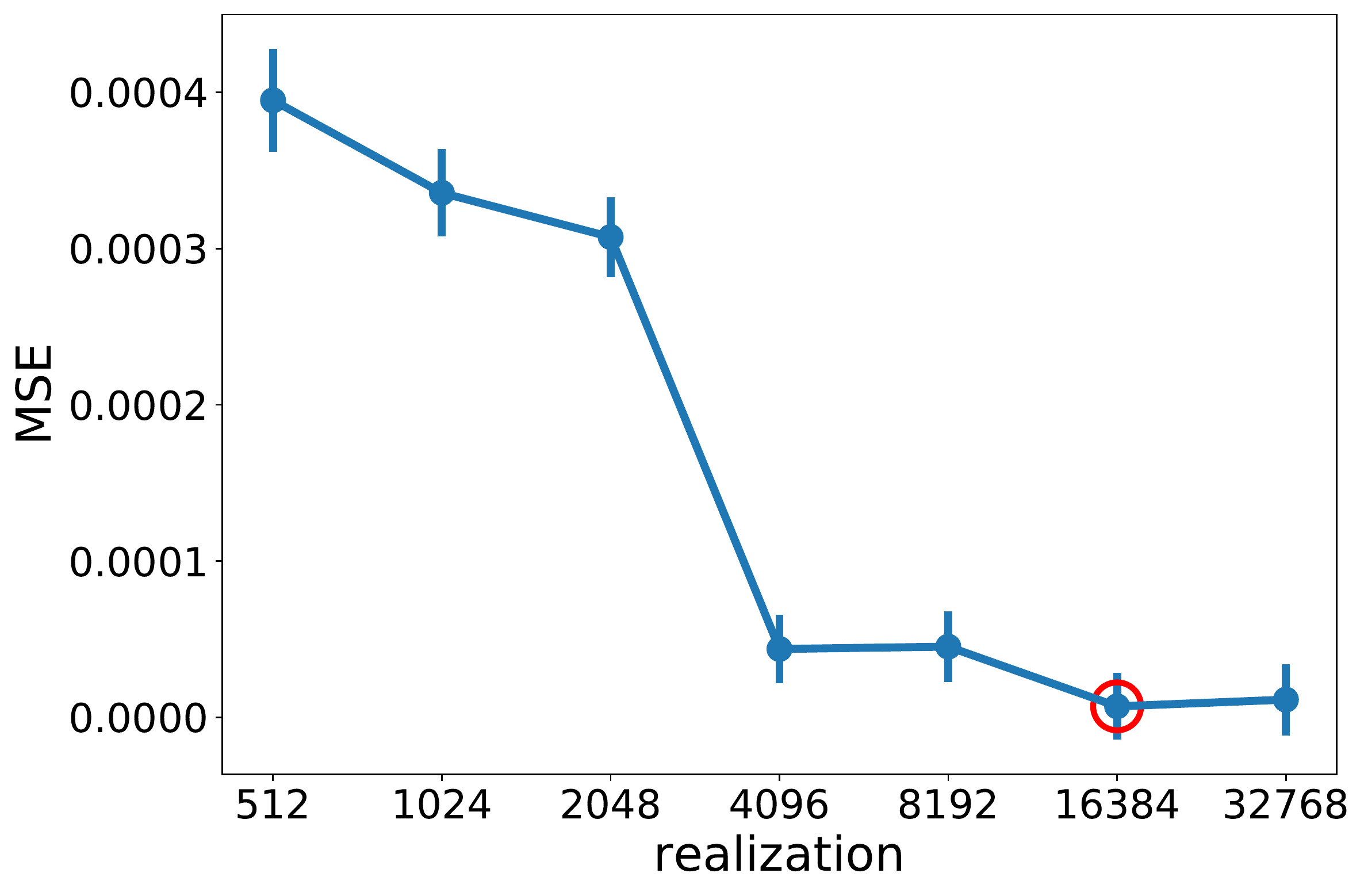}
\caption{The MSE vs. the number of realizations. We chose $N=16384$ as the number of realizations used in training.  Error bars represent 1 standard deviation from the result of 10 independent runs of the learning algorithm.}
\label{f:data_size}
\end{center}
\end{figure}

We systematically increased the number of realizations during the Monte-Carlo simulations to find a balance between the computation cost and model accuracy. Fig.~\ref{f:data_size} shows the MSE (at the 2000$^{th}$ epoch) against increasing numbers of realizations. Based on this information, we set the number of realizations generating the ensemble of examples to to $N=2^{14}=16384$. 
The number of realizations needed to fully cover the feature space is generally determined by some factors such as 1) the number of features in the space, and 2) the spanning width (domain) of the each feature.  As illustrated in Fig.~\ref{f:data_size}, $N=16384$ represents a good compromise between computation costs and minimizing the MSE.

\section{Testing the Learned Model}
\label{validating}

To test the accuracy of the learned model, we computed numerical ``experiments" on several tissues (brain and liver) that were of different scales, and whose geometries were significantly different than the simple unit cells illustrated in Fig.~\ref{f:RepGeo}. The goal of the validation was to compare 1) averaged concentrations computed by spatially filtering the direct microscale simulations on the complex geometries, and 2) averaged concentrations computed from the upscaled balance combined with the prediction of the effectiveness factor from the trained MLP network.  The validation process involved the following sequence of activities.
\begin{enumerate}
    \item Generation of realistic microscale test problems with appropriate parameters and tissue geometry in 2-dimensions (i.e., Table~\ref{T:range} and Fig.~\ref{f:cells}).
    \item Solving the test problems for specified values of the P\'eclet number, and spatially averaging them along the direction of the mean velocity to develop a 1-dimensional projection of the averaged concentration.
    \item Developing and solving the equivalent 1-dimensional representation of the upscaled equation given by Eq.~\eqref{one-eq}.  This equation contains a reaction rate term of the form given by Eq.~\eqref{effective_def}. This requires determining estimates of $\langle C_{\beta}\rangle^\beta$ and $\partial \langle C_{\beta}\rangle^\beta/\partial Z$ during the computation, and appropriately matching the P\'eclet number on the complex and training domains.
    \item Comparison of the average concentration transects computed directly from the averaged microscale equations, and from the upscaled equation using the learned effectiveness factor.
\end{enumerate}
These steps are described in the material following.

\subsection{Generating the Microscale Representations from Observed Tissue Structures}

On of the goals of this work is to illustrate that the learned model for the effectiveness factor is robust in the sense that it provides a direct macroscale prediction of the average concentration that is consistent with the concentration fields averaged from microscale simulations.  Good correspondence between these two measures indicates that the closure has been effectively carried out, and the micro-macro connection is well represented by the closure.  Toward that end, we computed the microscale solution for transport and reaction problems on two tissues whose structure was adapted from images reported in the literature.

The detailed 2-dimensional simulations for validation were based on the brain \citep{chen2000changes} and liver \citep{leedale2020multiscale} geometries illustrated in Fig.~\ref{f:cells}; the arrows denote the direction of mean flow (left to right) for these two domains.  
In order to assure that a large range of concentrations and concentration gradients were generated in our validation simulations, we extended the domain for both geometries using a sequence of reflections and concatenations.  First, each domain was reflected about the right-hand axis perpendicular to the mean flow direction to generate a 1 by 2 domain, periodic on the boundaries perpendicular to mean flow.  A second such reflection created a 1 by 4 domain, periodic at the inlet (left) and outlet (right).  Additionally, to create reasonably unconstrained boundaries in the direction perpendicular to flow, the entire set of 1 by 4 cells was reflected about the axis parallel to flow.
The final results where the creation of 4 by 2 arrays of the original geometries as shown in Fig.~\ref{f:validation}.

Solving the transport problem on these 2D domains was done using the finite element code COMSOL; the approach, and error metrics were as described in \S \ref{learning} 
The set of parameters used for the simulations of brain and liver is provided in Table~\ref{T:param}; these data are consistent with values determined from the literature listed in Table \ref{T:range}. 

\begin{table}[t]
\centering
    \caption{Parameters used for direct microscale simulations for brain and liver.}
    \vspace{2mm}
    \begin{tabular}{l|l|l}
        \hline
        ~~Parameter~~                                 & ~~~Brain~~~             & ~~~Liver~~~             \\
        \hline
        $r_{\sigma,\textrm{eff}}$ ($\mu m$)       & $0.47$            & $11.67$           \\
        $L_0$ ($\mu m$)                           & $5$               & $170$             \\
        $L$ ($\mu m$)                             & $20$              & $680$             \\
        $D_\beta$ ($\frac{m^2}{s}$)               & $1\times10^{-10}$ & $1\times10^{-10}$ \\
        $D_\sigma/D_\beta$                        & $0.1$             & $0.1$             \\
        $c_{\textrm{inlet}}$ ($\frac{mol}{m^3}$)  & $2$               & $2$               \\
        $\varepsilon_\beta$                       & $0.26$            & $0.187$           \\
        $k_m$ ($\frac{mol}{m^3\cdot s^{-1}}$)     & $226.34$          & $1.5$             \\
        $K$ $\frac{mol}{m^3}$                     & $1$               & $1$               \\
        $Pe$                                      & $8.16$            & $52.09$           \\
        $\varphi^2$                               & $0.5$             & $2.04$            \\
        $\kappa$                                  & $1\times10^{-10}$ & $1\times10^{-10}$ \\
        \hline
    \end{tabular}
    \label{T:param}
\end{table}

In order to compare the results obtained by the machine learned effectiveness factor with 2-D microscale simulation results (Eqs.~\eqref{microe}-\eqref{microi}), we calculated the averaged concentration at the cross-sectional direction in the flow phase for 2-D microscale simulations.  The values for $\langle C_\beta \rangle^\beta$ were determined using the average defined by Eq.~\eqref{volavebeta}, where the weighting function, $w$, was taken to be a uniform (top-hat) function with width equal to $2\,r_{\sigma,eff}$, and height equal to the domain height.

We present one computation example for the each different geometries (brain and liver) using physiologically reasonable parameter values (Table \ref{T:param}).  The green line in Fig.~\ref{f:validation}(A) illustrates the averaged concentration profile resulted from solving Eqs.~\eqref{mom}-\eqref{microi} in a 4 by 2 array of the brain geometry and associated parameters; the green line in Fig.~\ref{f:validation}(B) provides corresponding information for the liver geometry and associated parameters.

\subsection{The 1-Dimensional Averaged Equation}

Under steady-state conditions, with the average velocity aligned with the $z$-axis, our closed version of the macroscale transport equation Eq.~\eqref{one-eq} takes the form 
\begin{equation}
0 = - U  \frac{\partial \langle c_{\beta}\rangle^\beta}{\partial z} 
+D^* \frac{\partial^2 \langle c_{\beta}\rangle^\beta}{\partial z^2}
-{\eta}\frac{\varepsilon_\sigma}{\varepsilon_\beta} k_m \frac{\langle c_{\beta}\rangle^\beta}{\langle c_{\beta}\rangle^\beta+ K}
\label{ss}
\end{equation}
This represents the upscaled (coarse-grained) mass balance problem for the tissue.  The advantage to using the upscaled balance is that it requires significantly fewer degrees of freedom to resolve compared with the direct microscale solutions in 2-dimensions.

A custom-made steady-state finite difference code was generated to solve the macroscale transport problem directly; it uses the trained MLP for prediction of the appropriate value of the effectiveness factor, $\eta$.  For these simulations, the external boundaries were set as follows.

\begin{enumerate}
\item Inlet. Dirichlet boundary condition specifying the macroscale concentration.
\item Outlet.  Neuman boundary condition imposing a zero-gradient in the concentration.
\item Edges perpendicular to mean flow direction.  Periodic boundary conditions for the concentration.
\end{enumerate}

The solution algorithm used second-order centred differences for spatial derivatives; conventional Picard iteration was used to solve the system.  The stopping criterion for convergence was a global pointwise tolerance of $\epsilon = 5\times10^{-10}$ between successive iterations.  The value of $\eta$ was determined by calling the trained network prediction once every ten Picard iterations for the purpose of saving computation time.  The value of $\eta$ was updated using the current value of both explicit and implicit features ($\varepsilon_\beta, Pe, \phi^2, D_r, \langle C_\beta\rangle^\beta,$ and $\tfrac{\partial}{\partial Z} \langle C_\beta\rangle^\beta$); note that only the implicit features must be updated during computation).   The grid convergence analysis was performed to ensure that the the GCI is of order $1 \times10^{-7}$ \citep{roache1994perspective}.

In addition to the effectiveness factor, the upscaled model also requires the effective dispersion coefficient, $D^*$.   We computed $D^*$ using the conventional methods of volume averaging \citep{whitaker1999,wood2011dispersive}.  For completeness, the approach is described in detail in the Appendix.   

In order to compare the microscale and macroscale computations, it was necessary to match the values of the P\'eclet number ($Pe$) and the Thiele modulus ($\varphi^2$) between the microscale and macroscale simulations.  Each of these unitless numbers require a characteristic length estimate for the effective cell radius $r_{\sigma,eff}$.  

While $r_{\sigma,eff}$ is easy to estimate for the simple learning domain illustrated in Fig.~\ref{f:RepGeo}, it is challenging to determine it for the complex geometries illustrated in Fig.~\ref{f:cells}. 
This is not a trivial issue; the accurate values of both $Pe$ and $\varphi^2$ (both depend on $r_{\sigma,eff}$) are needed as input to the trained network in order to predicting $\eta$ and, consequently, the effective reaction rate with high fidelity.  Recall, $Pe$ and $\varphi^2$ are defined by
\begin{align}
Pe &= \frac{U~r_{\sigma,eff}}{D_\beta}
\label{Pe_eff} \\
\varphi^2 &= \frac{ k_m r^2_{\sigma,eff}}{ K \mathscr{D}_{\beta}}
\end{align}
Note that each of these parameters requires an estimate of an \emph{effective} value for the cell radius; we denote this quantity by $r_{\sigma,eff}$. While each of the other parameters comprising $Pe$ and $\varphi^2$ are set independently of the geometry, $r_{\sigma,eff}$ must come from the geometry itself.  The challenge at this juncture is to develop a reasonable method for predicting an effective radius, $r_{\sigma,eff}$, for the complex geometries illustrated in Fig.~\ref{f:RepGeo}.  

Following previous work \citep{ostvar2016}, we used a Voronoi decomposition method to determine this value. 
Voronoi tessellation provides one method to assign an effective radius to complex geometries that are not inherently circular.  The method is known generally as Voronoi decomposition; the approach is well documented elsewhere \citep[e.g.,][\S 17.4]{skiena2008}. As a summary, given a finite set of points in a plane, for each point the corresponding convex polygons Voronoi cell is formed by all the locations closer to that point than to any of the other points.  To compute the effective radius of the tessellated domain, maximum-sized circles were inscribed in each Voronoi cell. The effective radius, $r_{\sigma,eff}$, was taken as the \emph{area weighted} radius of these inscribed circles. The resulting values for brain and liver were computed to be $r_{\sigma,eff}=0.47$ and $11.67~\mu m$, respectively.  With these values established, it was possible to compute $Pe$ and $\varphi^2$ uniquely for each of the two domains. This allowed matching both the P\'eclet number and Thiele modulus, which are necessary parameters for computing $D^*$ and $\eta$.

\subsection{Comparison of the Results}

\begin{figure*}[t]
\begin{center}
\includegraphics[width=.9\textwidth]{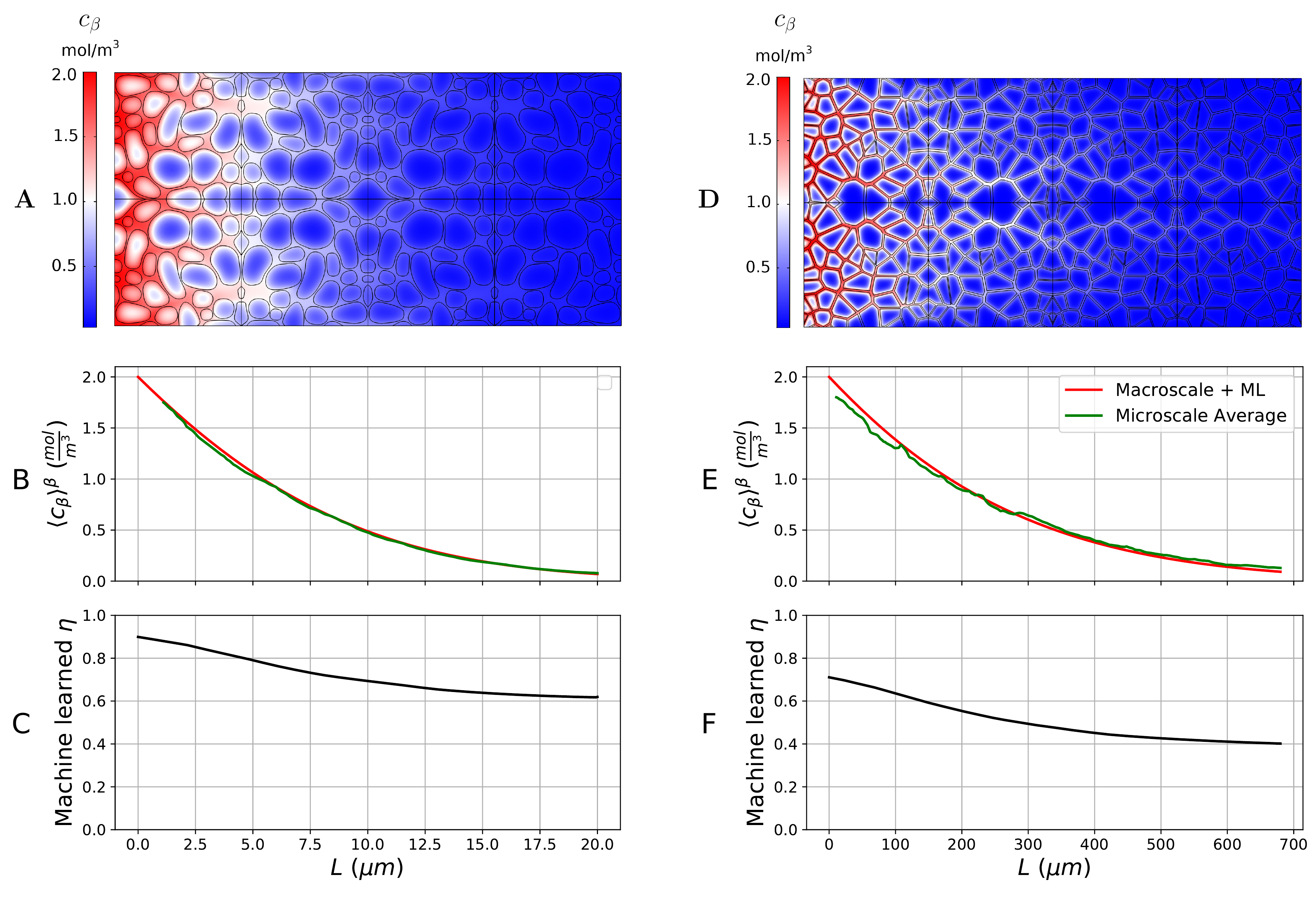}
\caption{(A) and (D) The concentration profiles resulting from solving Eqs.~\eqref{mom}-\eqref{microi} in a 4 by 2 array of the brain and liver geometries. (B) and (C) The comparison of the vertically averaged concentrations from the microscale numerical (blue) and upscaled (red) equations.  
(C) and (F) The learned effectiveness factors as a function of distance along the direction of flow. }
\label{f:validation}
\end{center}
\end{figure*}

Figure~\ref{f:validation} shows a comparison of 1) the averaged concentration calculated from the averaged microscale concentration profile, and 2) the concentration obtained by solving the coarse-grained mass balance (Eq.~\eqref{ss}) coupled with the trained network for predicting $\eta$.  Fig~\ref{f:validation}(C) shows the corresponding effectiveness factor computed by the trained network.  There is generally a good correspondence between the two averaged concentrations. Note that the geometries for the brain and liver systems have dramatically different length scales ($L_0=5$ and $L_0=170~\mu m$), respectively. Interestingly, the learned model is able to capture the 1-D upscaled concentration regardless of the difference in their length scales. This is, in part, because we have used the dimensionless feature set to train the model. Even though the model was trained on a resembled geometry in which all the necessary data for training were collected within that geometry, it can predict the upscaled concentrations.

We defined the following error metric for the averaged spatial concentration.  First, the average concentration computed from the 2-dimensional microscale solutions were interpolated linearly to the same grid used for the finite-difference computation of Eq.~\eqref{ss}.  This provided 1) the average concentration predicted from the 2-dimensional microscale simulations, and 2) the average concentration predicted from the macroscale balance given in Eq.~\eqref{ss} on a uniform, 1-dimensional grid with $N$ total grid points. The $L^1$ error metric for each point was then computed from

\begin{align}
\epsilon(z_i)=\frac
{\left|\langle c_\beta \rangle^\beta_{\textrm{macro+ML}}-\langle c_\beta  \rangle^\beta_{\textrm{micro}}\right|}
{\langle c_\beta \rangle^\beta_{\textrm{macro+ML}}}
\label{err}
\end{align}
and the average fractional error was then computed from 

\begin{equation}
\epsilon_{avg} = \frac{1}{N} \sum_{i=i}^{i=N} \epsilon(z_i)\\
\end{equation}
This result computes the average error which are 2.6\% and 9.6\% respectively.

\section{Conclusions}

In this work, we were able to provide a novel data-driven approach to \emph{closure} for upscaling nonlinear problems.  In particular, for the problem of upscaling nonlinear, Michaelis-Menten-type kinetics in tissues, we were able to present a workflow structure that defined the upscaling process, and the subsequent development and training of a MLP network for closing the upscaled problem.  The MLP network was trained on a model with particularly simple geometry; however, this did not appreciably limit the accuracy of the trained network for making predictions in more complex, but nearly isotropic, geometries.  This result is consistent with previous observations in upscaled media.  The result is significant in that it suggests for nearly isotropic tissue structures, the exact geometry of the tissue (e.g., spherical versus polygonal) is not as important as more basic metrics such as volume fraction.  The practicality of this observation is that the trained networks exhibit high generalizability.  This strongly suggests (as validated by our results), that a single trained network can be effectively applied for an entire class of problems defined by the form of the nondimensionalized equations, and the rough geometrical attributes.  For investigations where an explicit form for the upscaled balance equation is desired, the approach reported here provides one possible route for providing a closed and explicit macroscale balance equation. More importantly, we showed that the implicit features derived from the source term appearing in the microscale closure problem are essential for the network in order to predict the correction factor with high fidelity; this can be attributed to the preserving the mass balance in the system for the case where these physics-driven features are included in the feature set. In other words, incorporating the information from both microscale (explicit features) and macroscale (implicit features) significantly boosts the neural network accuracy.

\subsection*{Data and Code Availability}
All data, finite different, and deep learning codes used in this paper are available on \url{https://github.com/TaghizadehE/implicit-PINN}

\section{Acknowledgment}
This work was supported by the School of Chemical, Biological and Environmental Engineering at Oregon State University through the Graduate Teaching Assistantship given to ET. BDW acknowledges Steve Whitaker for collaborative work on this topic started several decades ago.  We are grateful to Dr. Joseph Leedale (Univ. Liverpool) for graciously sharing his segmented geometries for the hepatocyte sphereoids.

\appendix
\section{~}

\subsection{Theoretical Calculations of the Effective Dispersion}

In 1-D, the macroscale dispersion tensor is represented by two independent parts
\begin{align}
    {D}^*_{zz} &= D_{\textit{eff}} - \langle \tilde{v}_z b_z \rangle^\beta
\label{deff}    
\end{align}
The effective dispersion tensor was determined using conventional volume averaging methods \citep[][Chp. 3]{whitaker1999}. The value of $b_z$ is determined from the following boundary value problem solved over the representative domain (unit cell) described previously by \citep{wood2011dispersive}

\begin{align}
    \tilde{v}_x +{\bf v}_\beta \cdot \nabla b_z &= \mathscr{D}_\beta \nabla^2 b_z\\
    -{\bf n}_{\beta\sigma}\cdot \mathscr{D}_\beta \nabla b_z &= n_z \mathscr{D}_\beta \\
    b_z({\bf r}+\mathbf{\ell}_i) &= b_z({\bf r}) \\
    \langle b_z \rangle &= 0
\label{b-field}
\end{align}
This problem was solved numerically using the same code and approach as described for the solution to the microscale balance equations.  The result of this computation is the $b_z$ field, from which the effective dispersion coefficient $D^*_{zz}$ is computed by taking the average of the solution as shown in Eq.~\eqref{deff}.

To compute the effective diffusion coefficient, we adopted the solution described by \citep{ochoa1994diffusive}.

\begin{equation}
    \frac{D_{\textit{eff}}}{\mathscr{D}_\beta} = 
    \frac{2 \mathscr{D}_\sigma/\mathscr{D}_\beta-\varepsilon_\beta(\mathscr{D}_\sigma/\mathscr{D}_\beta-1)}{2+\varepsilon_\beta(\mathscr{D}_\sigma/\mathscr{D}_\beta-1)}
\end{equation}
where $\mathscr{D}_\sigma$ and $\mathscr{D}_\beta$ are extracellular and intercellular diffusion coefficient, respectively. The effective diffusion coefficient can also be computed slightly more accurately by solving a \emph{closure problem} as described in \citep[][Chp.~1]{whitaker1999}.  Because the diffusion coefficient is not strongly affected by geometry for nearly isotropic materials, the analytical approximation was determined to be sufficiently accurate. The normalized (by $\mathscr{D}_\beta$) effective diffusion coefficient is depicted in Fig.~S2.

\begin{figure}
\begin{center}
\includegraphics[width=0.6\textwidth]{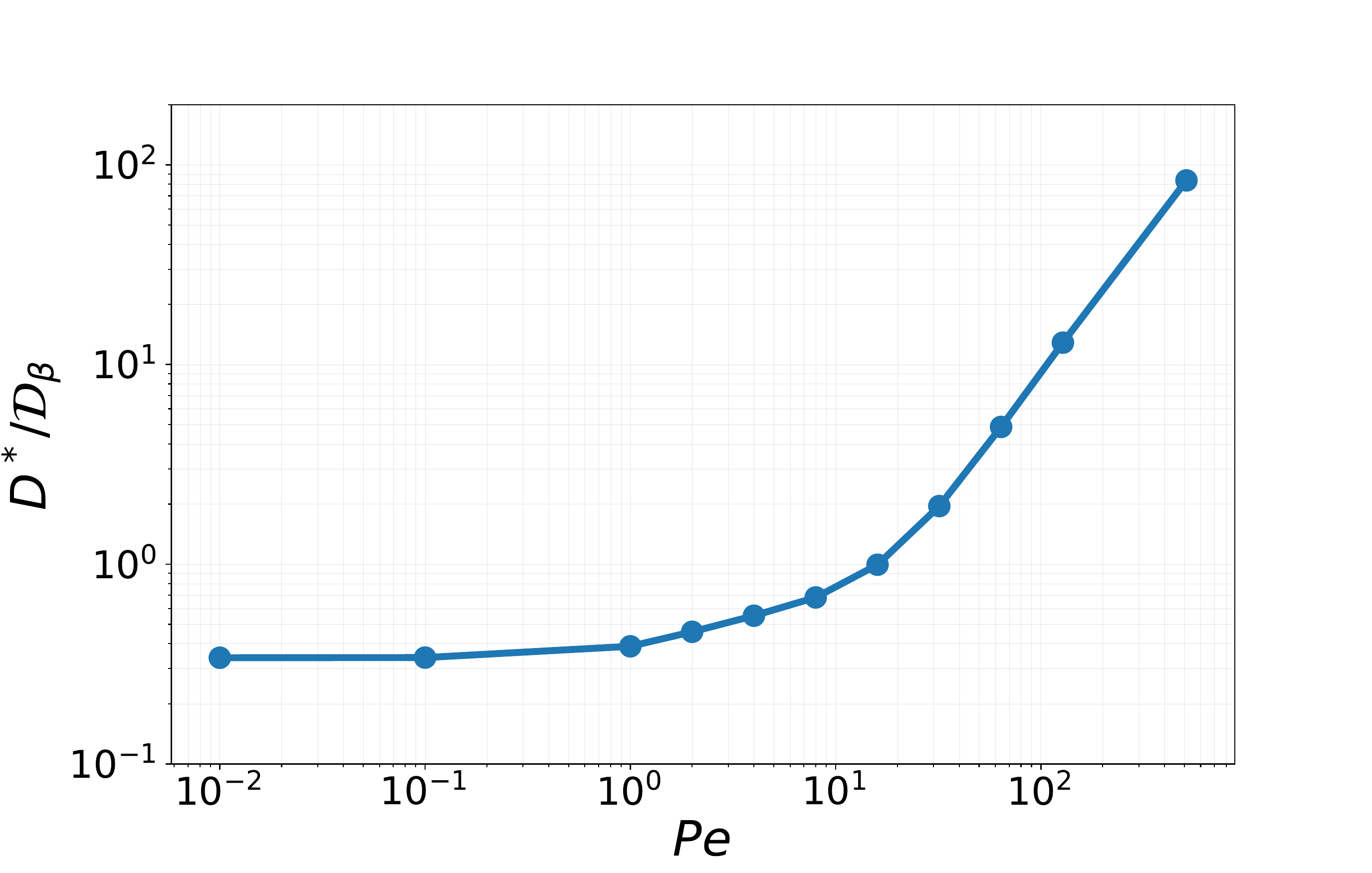}
\caption{Theoretical calculations of the effective dispersion vs. P{\'e} for the representative unit cell.}
\label{f:app:D_eff}
\end{center}
\end{figure}





\subsection{Simulation Work Flow}
Here we present the simulation work flow for the brain geometry. It is analogous for the liver case. The simulation domain was set up containing five square unit cells, each with a side length of 3 $\mu$m.  The domains were set up to be geometrically similar to brain tissue geometries reported in the literature, although this choice was arbitrary in that any similarly structured tissue would yield equivalent results.  We established the following variables to use as features in combination for predicting the effectiveness factor.     
\begin{enumerate}
    \item Extracellular phase volume fraction, $\varepsilon_\beta$
    \item Thiele modulus, $\phi^2$
    \item P{\'e}clet number, $Pe$
    \item The ratio of intercellular to extracellular diffusion coefficients, $D_r$
    \item Normalized spatial average concentration in the $\beta$ phase, $\langle C_\beta \rangle^\beta$
    \item Gradient of the spatial average concentration $\beta$ phase, $\frac{\partial \langle C_\beta \rangle^\beta}{\partial X}$
\end{enumerate}

For these simulations, the value of the microscale extracellular diffusion coefficient was fixed at $\mathscr{D}_{\beta} = 1\times 10^{-10}~(m^2/s)$ and half-saturation constant was fixed at $K=1~(mol/m^3)$.

For computations, each simulation was run in dimensional form. The associated values of $\varepsilon_\beta$, $Pe$, $\varphi^2$, and $D_r$ were computed from the data randomly selected for each realization.  The parameter selection process can be outlined as follows.

\begin{enumerate}
    \item Choose the radius of the particles $0.5\times 10^{-6} < r_{\sigma,eff} < 1.02\times 10^{-6}~(m)$; this makes the $0.25<\varepsilon_\beta<0.85$. 
    \item  Choose $\varphi^2$ ($1\times 10^{-2} < \varphi^2 < 1\times 10^{4}$). Chosen with a uniform random number; then rescale and change the distribution by
    $\varphi^2 = 100 - \sqrt{\varphi^2}$. This should be equivalent to $\varphi^2$ from 0.01 to 100.
    
    \item $p(x^-)$ from 0.1 to $3\times 10^{6}~(Pa)$, chosen randomly; then rescale and change the distribution. This is the value to use at the left boundary. $p(x^+) = 1\times 10^{5}~(Pa)$ is the fixed value for the right boundary. This should be equivalent to P{\'eclet} numbers between 0.1 and 100.
    
    \item $D_r$ from 0.01 to 1, chosen randomly. Set $\mathscr{D}_\beta$ equal to $1\times 10^{-10}~({m^2}/{s})$. Compute the $\mathscr{D}_\sigma = D_r~\mathscr{D}_\beta$.
    
    \item $c_{\beta}(x^-)$ from 0.1 to 10 $(mol/m^{3})$. Chosen randomly.  This is the value to use at the left boundary. The concentration at the right was then set to the zero normal derivatives, or continuative boundary.    
    
\end{enumerate}

Note that ${\bf x}=(x,y,z)$, and the flow is in the $x-$direction.  The unit cells are periodic in the other two directions.  Note that for the flow computation, the top and bottom boundaries can be set to zero tangential stress.  For the mass computation, the top and bottom boundaries can be set to no flux conditions (since there can be no flux because of periodic symmetry).

The sequence of computations would proceed as follows.  Each of the three unit cells has a total volume $V$, a fluid volume $V_\beta$, a porosity $\varepsilon_\beta = V_\beta/V$, and an area per unit volume equal to $a_v = A_{\beta\sigma}/V$. 

\begin{enumerate}
    \item Compute the $\varepsilon_\sigma$
    \item Compute the random values of the pressure boundaries ($p(x^-)$), compute the velocity field.
    \item Compute the value of $U$ (average velocity).
    \item Compute the P{\'e}clet number
    \item For this particular configuration for the flow field, generate $N$ random realizations of the transport and reaction problem.
        \begin{enumerate}
            \item Select the random value for $\varphi^2$.
            \item Compute the associated value of $k_m$ for recording later.
            \item Compute the \emph{steady state solution} for the concentration field for this configuration.
            \item Compute the intrinsic average concentration in each of the five unit cells, $\langle c_{\beta} \rangle^\beta_i$ (where $i=1,2,3,4$).
            \item Compute estimates of the gradient of the intrinsic average concentration for each of the five cells using a forward finite difference for cells 1, a centered finite difference for cell 2-4, and a backward finite difference for cell 5 \label{bob}.
            \item Compute the average of the $x-$derivative of the concentration field, $\langle \partial c_{\beta}/\partial x  \rangle^\beta$.  Note that this is, in general, different from the value computed in item \ref{bob} above.
            %
            %
            \item Compute the area average of the diffusive flux normal to the fluid-solid surface in \emph{each of the five unit cells}. 
            This gives the total (normalized) reaction rate in each of the five unit cells. Multiply this value by the area per unit volume divided by porosity ($a_v/\varepsilon_\beta$). Call this result $\langle R \rangle$ (with units mol$\cdot$m$^{-3}$s$^{-1}$).
            \item Compute the value of 
            \begin{equation}
                R_{0} = -\varepsilon_\sigma{k_m} \frac{\langle c_{\beta}\rangle^\beta} { \langle c_{\beta}\rangle^\beta +K}
            \end{equation}
            \item Compute the value 
                \begin{equation}
                    \eta = \frac{\langle R \rangle}{R_{0}}
                \end{equation}
        \item Because there are essentially five unit cells in every simulation, each simulation that is run will generate four dataset. Record the following parameters 
            \begin{enumerate} 
                \item variables: $\varepsilon_\sigma$, $\varphi^2$, $Pe$, $D_r$, $\eta$
                \item variables: $\langle c_{\beta}\rangle^\beta$, $\tfrac{\partial}{\partial x} \langle c_{\beta}\rangle^\beta$, $\langle \tfrac{\partial}{\partial x}c_{\beta}\rangle^\beta$
                \item Ancillary variables: $U$, $p(x^-)$, $k_m$
                \item variable: $\langle R \rangle$, $R_0$
            \end{enumerate}
         \item Go back to item 4(a) until $N$ realizations are completed
         \end{enumerate}
    \item Go back to item 1 to compute a new velocity field.

\end{enumerate}

\bibliography{main}

\end{document}